\documentclass[aos,MSNbibl,nameyear,seceqn,dvips]{arximspdf}
\usepackage{graphicx}

\doi{10.1214/11-AOS901}
\volume{39}
\issue{4}
\pubyear{2011}
\firstpage{2205}
\lastpage{2242}

\makeatletter

\newtheorem{theorem}{Theorem}[section]
\newtheorem{prop}[theorem]{Proposition}
\newtheorem{lem}[theorem]{Lemma}
\newtheorem{cor}[theorem]{Corollary}

\newproclaim{rem}[theorem]{Remark}

\newcommand{\E}{\mathrm{E}}
\newcommand{\Ree}{\operatorname{Re}}
\newcommand{\Var}{\operatorname{Var}}
\newcommand{\Cov}{\operatorname{Cov}}
\newcommand{\ii}{\mathrm{i}}
\newcommand{\argmin}{\arg\min}

\makeatother

\begin{document}
\begin{frontmatter}

\title{Statistical inference for time-changed L\'evy processes via composite characteristic function estimation}
\runtitle{Statistical inference for time-changed L\'evy processes}

\begin{aug}
\author[A]{\fnms{Denis} \snm{Belomestny}\corref{}\ead[label=e1]{denis.belomestny@uni-due.de}}
\runauthor{D. Belomestny}
\affiliation{Duisburg-Essen University}
\address[A]{Duisburg-Essen University\\
Forsthausweg 2, D-47057 Duisburg\\
Germany\\
\printead{e1}} 
\end{aug}

\thankstext{t1}{Supported in part by SFB 649 ``Economic Risk.''}

\received{\smonth{12} \syear{2010}}
\revised{\smonth{4} \syear{2011}}

%
\begin{abstract}
In this article, the problem of semi-parametric inference on the
parameters of a multidimensional L\'evy process $ L_{t} $ with
independent components based on the low-frequency observations of the
corresponding time-changed L\'evy process $ L_{\mathcal{T}(t)}$,
where $ \mathcal{T} $ is a nonnegative, nondecreasing real-valued
process independent of $ L_{t}$, is studied. We show that this
problem is closely related to the problem of composite function
estimation that has recently gotten much attention in statistical
literature. Under suitable identifiability conditions, we propose a
consistent estimate for the L\'evy density of $ L_{t} $ and derive
the uniform as well as the pointwise convergence rates of the
estimate proposed. Moreover, we prove that the rates obtained are
optimal in a minimax sense over suitable classes of time-changed L\'evy
models. Finally, we present a simulation study showing the performance
of our estimation algorithm in the case of time-changed Normal Inverse
Gaussian (NIG) L\'evy processes.
\end{abstract}

%

\pdfkeywords{62F10, 62J12, 62F25, 62H12, Time-changed Levy processes,
dependence, pointwise and uniform rates of convergence,
composite function estimation}

\end{frontmatter}

\section{Introduction}
The problem of nonparametric statistical inference for jump processes
or more generally for semimartingale models has long history and goes
back to the works of \citet{RT} and \citet{BB}. In the past decade,
one has witnessed the revival of interest in this topic which is
mainly related to a wide availability of financial and economical time
series data and new types of statistical issues that have not been
addressed before. There are two major strands of recent literature
dealing with statistical inference for semimartingale models. The first
type of literature considers the so-called high-frequency setup, where
the asymptotic properties of the corresponding estimates are studied
under the assumption that the frequency of observations tends to
infinity. In the second strand of literature, the frequency of
observations is assumed to be fixed (the so-called low-frequency setup)
and the asymptotic analysis is done under the premiss that the
observational horizon tends to infinity. It is clear that none of
the above asymptotic hypothesis can be perfectly realized on real
data and they can only serve as a convenient approximation, as in
practice the frequency of observations and the horizon are always
finite. The present paper studies the problem of statistical inference
for a class of semimartingale models in low-frequency setup.

Let $X = (X_t)_{t\geq0} $ be a stochastic process valued in $ \mathbb
{R}^{d} $ and let $\mathcal{T} = (\mathcal{T}(s))_{s\geq
0} $ be a nonnegative, nondecreasing stochastic process not necessarily
independent of $ X $ with $ \mathcal{T}(0)=0 $. A time-changed process
$Y = (Y_s)_{s\geq0} $ is then
defined as $Y_s = X_{\mathcal{T}(s)}$. The process $\mathcal{T} $ is
usually referred to as time
change. Even in the case of the one-dimensional Brownian motion $ X$,
the class of time-changed processes $ X_{\mathcal{T}} $ is very large
and basically coincides with the class of all semimartingales [see,
e.g., \citet{M}]. In fact, the construction in
\citet{M} is not direct, meaning that the problem of
specification of
different models with the specific properties remains an important
issue. For example, the base process $X $ can be assumed to possess
some independence property
(e.g., $ X $ may have independent components), whereas a nonlinear
time change can induce deviations from the independence. Along this
line, the time change can be used to model dependence for stochastic
processes. In this work, we
restrict our attention to the case of time-changed L\'evy processes,
that is,
the case where $X=L$ is a multivariate L\'evy process and
$\mathcal{T} $ is an independent of $ L $ time change.
Time-changed L\'evy processes are one step further in increasing
the complexity of models in order to incorporate the so-called stylized features
of the financial time series, like volatility clustering [for more
details, see \citet{CGMY}]. This type of processes
in the case of the one-dimensional Brownian motion was first studied by
\citet{Bo}. \citet{Cl} introduced Bochner's time-changed
Brownian motion into financial economics: he used it to relate future
price returns of cotton to the variations in volume during
different trading periods.
Recently, a number of parametric time-changed L\'evy processes
have been introduced by \citet{CGMY}, who model the stock price $
S_{t} $ by a geometric time-changed L\'evy model
\[
S_{t}=S_{0}\exp\bigl(L_{{\mathcal{T}(t)}}\bigr),
\]
where $ L $ is a L\'evy process and $ \mathcal{T}(t) $ is a time change
of the form
%
%
\begin{equation}
\label{TCD}
\mathcal{T}(t)=\int_{0}^{t}\rho(u) \,du
\end{equation}
with $ \{ \rho(u) \}_{u\geq0} $ being a positive mean-reverting process.
\citet{CGMY} proposed to model $ \rho(u) $ via the
Cox--Ingersoll--Ross (CIR) process.
Taking different parametric L\'evy models for $ L $ (such as
the normal inverse Gaussian or the variance Gamma processes) results in
a wide range of processes
with rather rich volatility structure (depending on the rate process
$\rho$) and various distributional properties (depending on the
specification of $ L $).
From statistical point of view, any parametric model (especially one
using only few parameters) is prone to misspecification problems.
One approach to deal with the
misspecification issue is to adopt the general nonparametric models for
the functional
parameters of the underlying process. This may reduce the estimation
bias resulting
from an inadequate parametric model.
In the case of time-changed L\'evy models, there are two natural
nonparametric parameters: L\'evy density $ \nu$, which determines the
jump dynamics of the process $ L $ and the marginal distribution of the
process $ \mathcal{T}$.

In this paper, we study the problem of statistical inference on the
characteristics of a multivariate L\'evy process $ L $ with independent
components based on low-frequency observations of the time-changed
process $ Y_{t}=L_{\mathcal{T}(t)}$, where $ \mathcal{T}(t) $ is a
time change process independent of $L$ with strictly stationary
increments. We assume that the distribution of $ \mathcal{T}(t) $
is unknown, except of its mean value.
This problem is rather challenging  and has not been yet given
attention in the literature, except for the special case of $ \mathcal
{T}(t)\equiv t $ [see, e.g., \citet{NR} and \citet{CG}].
In particular, the main difficulty in constructing nonparametric
estimates for the L\'evy density $ \nu$ of $ L $
lies in the fact that the jumps are unobservable variables, since in practice
only discrete observations of the process $ Y $ are available. The more
frequent  the observations, the more relevant information about
the jumps of the underlying process,
and hence, about the L\'evy density $ \nu$ are contained in the
sample. Such high-frequency based statistical
approach has played a central role in the recent literature on
nonparametric estimation
for L\'evy type processes. For instance, under discrete observations of
a pure L\'evy process
$ L_{t} $ at times $ t_{j}=j\Delta, j=0,\ldots, n$,
\citet{W} and \citet{FL1}
proposed the quantity
\[
\widehat\beta(f)=\frac{1}{n\Delta}\sum_{k=1}^{n} f (L_{t_{k}}-L_{t_{k-1}})
\]
as a consistent estimator for the functional
\[
\beta(f)=\int f (x)\nu(x) \,dx,
\]
where $ f $ is a given ``test function.''
Turning back to the time-changed L\'evy processes, it was shown in
\citet{FL2} [see also \citet{RTA}] that in the case, where
the rate process $ \rho$ in (\ref{TCD}) is a positive
ergodic diffusion independent of the L\'evy process $ L $, $ \widehat
\beta(f) $ is still a consistent
estimator for $ \beta(f) $ up to a constant, provided the time horizon
$ n \Delta$ and
the sampling frequency $ \Delta^{-1} $ converge to infinite at
suitable rates.
In the case of low-frequency data ($ \Delta$ is fixed), we cannot be
sure to what extent the
increment $ L_{t_{k}}-L_{t_{k-1}} $ is due to one or several jumps or
just to the
diffusion part of the L\'evy process so that at first sight it may
appear surprising that some kind of inference in this situation is
possible at all. The key observation here is that for any bounded
``test function'' $f $
%
%
\begin{equation}
\label{CONVERG}
\frac{1}{n}\sum_{j=1}^{n}f\bigl(L_{\mathcal{T}(t_{j})}-L_{\mathcal
{T}(t_{j-1})}\bigr)\to\E_{\pi}\bigl[f\bigl(L_{\mathcal{T}(\Delta)}\bigr)\bigr],\qquad
n\to\infty,
\end{equation}
provided the sequence $\mathcal{T}(t_{j})-\mathcal{T}(t_{j-1}),
j=1,\ldots, n$, is stationary and ergodic with the invariant stationary
distribution $ \pi$.
The limiting expectation in~(\ref{CONVERG}) is then given by
\[
\E_{\pi}\bigl[f\bigl(L_{\mathcal{T}(\Delta)}\bigr)\bigr]
=\int_{0}^{\infty} \E
[f(L_{s})]
\pi(ds).
\]
Taking
$ f(z) = f_{u}(z)=\exp(\ii u^{\top}z), u\in\mathbb{R}^{d}$, and using
the independence of $L$ and~$\mathcal{T}$, we arrive at the following
representation for the c.f. of $ L_{\mathcal{T}(s)} $:
%
%
\begin{equation}
\label{EstEq}
\E\bigl[\exp\bigl(\ii u^{\top}L_{\mathcal{T}(\Delta)}
\bigr)\bigr]=\int
_{0}^{\infty} \exp(t\psi(u)) \pi(dt)=\mathcal{L}_{\Delta}(-\psi(u)),
\end{equation}
where $ \psi(u)=t^{-1}\log[\E\exp(\ii u L_{t})] $ is the characteristic
exponent of the L\'evy process $ L $ and $ \mathcal{L}_{\Delta} $
is the Laplace transform of $ \pi$. In fact, the most difficult part
of estimation procedure comes only now and consists in reconstructing
the characteristics of the underlying L\'evy process $ L $ from an
estimate for $ \mathcal{L}_{\Delta}(-\psi(u))$. As we will see, the
latter statistical problem is closely related to the problem of
composite function estimation, which is known to be highly nonlinear
and ill-posed. The identity (\ref{EstEq}) also reveals the major
difference between high-frequency and low-frequency setups. While in
the case of high-frequency data one can directly estimate linear
functionals of the L\'evy measure $ \nu$, under low-frequency
observations, one has to deal with nonlinear functionals of $ \nu$
rendering the underlying estimation problem nonlinear and ill-posed.
 Last but not  least, the increments of time-changed L\'evy
processes are not any longer independent, hence advanced tools from
time series analysis have to be used for the estimation of $ \mathcal
{L}_{\Delta}(-\psi(u)) $.

The paper is organized as follows. In Section \ref{TCL}, we introduce
the main object of our study, the time-changed L\'evy processes. In
Section \ref{SP}, our statistical problem is formulated and its
connection to the problem of composite function estimation is
established. In Section \ref{SA}, we impose some restrictions on the
structure of the time-changed L\'evy processes in order to ensure the
identifiability and avoid the ``curse of dimensionality.'' Section \ref
{EST} contains
the main estimation procedure. In Section \ref{ASYMP}, asymptotic
properties of the estimates defined
in Section \ref{EST} are studied. In particular, we derive uniform and
pointwise rates of convergence (Sections \ref{URC} and \ref
{PCR}, resp.) and prove their
optimality over suitable classes of time-changed L\'evy models
(Section~\ref{LBOUNDS}). Section~\ref{DISC} contains some discussion.
Finally, in Section~\ref{SIM} we present a~simulation study. The rest
of the paper contains proofs of the main results and some auxiliary
lemmas. In particular, in Section \ref{EXPBOUNDS} a useful inequality
on the probability of large deviations for empirical processes in
uniform metric for the case of weakly dependent random variables can be
found.\looseness=-1

\section{Main setup}
\subsection{\texorpdfstring{Time-changed L\'evy processes}{Time-changed Levy processes}}
\label{TCL}
Let $ L_{t} $ be a $ d $-dimensional L\'evy process on the probability
space $(\Omega,\mathcal{F},\mathbb{P})$ with the characteristic
exponent $
\psi
(u)$, that is,
\[
\psi(u)=t^{-1}\log\E[\exp(\ii u^{\top} L_{t}
)].
\]
We know by the L\'evy--Khintchine formula that
%
%
\begin{equation}
\label{psi}
\psi(u)=\ii\mu^{\top} u-\frac{1}{2}u^{\top}\Sigma u+
\int_{\mathbb{R}^{d}}\bigl( e^{\ii u^{\top}y}-1-\ii u^{\top}y\cdot
\mathbf
{1}_{\{ |y|\leq1 \}} \bigr)\nu(dy),
\end{equation}
where $ \mu\in\mathbb{R}^{d}, \Sigma$ is a positive-semidefinite
symmetric $ d\times d $ matrix
and $ \nu$ is a L\'evy measure on $ \mathbb{R}^{d}\setminus\{ 0 \} $
satisfying
\[
\int_{\mathbb{R}^{d}\setminus\{ 0 \}}(|y|^{2}\wedge1)\nu
(dy)<\infty.
\]
A triplet $ (\mu,\Sigma,\nu) $ is usually called a characteristic
triplet of the $ d $-dimensional L\'evy process $ L_{t}$.

Let $ t\to\mathcal{T}(t), t\geq0 $ be an increasing right-continuous
process with left limits such that $ \mathcal{T}(0)=0 $ and for each
fixed $ t $, the random variable $ \mathcal{T}(t) $ is a stopping time
with respect to the filtration $ \mathcal{F} $. Suppose furthermore
that~$\mathcal{T}(t)$ is finite
$ \mathbb{P}$-a.s. for all $ t\geq0 $ and that $ \mathcal{T}(t)\to\infty$
as $ t\to\infty$. Then the family of $ ( \mathcal{T}(t) )_{t\geq0} $
defines a random
time change.
Now consider a $ d $-dimensional process $ Y_{t}:=L_{\mathcal{T}(t)} $.
The process $ Y_{t} $ is called the time-changed L\'evy process. Let us
look at some examples. If $ \mathcal{T}(t) $ is a L\'evy
process, then~$ Y_{t}$ would be another L\'evy process. A more general
situation
is when $ \mathcal{T}(t) $ is modeled by a nondecreasing semimartingale
\[
\mathcal{T}(t)=b_{t}+\int_{0}^{t}\int_{0}^{\infty} y\rho(dy,ds),
\]
where $ b $ is a drift and $ \rho$ is the counting measure of jumps in
the time
change. As in \citet{CW}, one can take $ b_{t}=0 $ and consider
locally deterministic time changes
%
%
\begin{equation}
\label{TCProcess}
\mathcal{T}(t)=\int_{0}^{t}\rho(s_{-}) \,ds,
\end{equation}
where $ \rho$ is the instantaneous activity rate which is assumed to
be nonnegative.
When $ L_{t} $ is the Brownian motion and $ \rho$ is proportional\vadjust{\goodbreak}
to the instantaneous
variance rate of the Brownian motion, then $ Y_{t} $ is a pure jump L\'
evy process with
the L\'evy measure proportional to $ \rho$.
Let us now compute the characteristic function of $ Y_{t} $.
Since $ \mathcal{T}(t) $ and $ L_{t} $ are independent, we get
%
%
\begin{equation}
\label{CFY}
\phi_{Y}(u|t)=\E\bigl(e^{\ii u^{\top}L_{\mathcal{T}(t)}}\bigr)
=\mathcal{L}_{t}(-\psi(u)),
\end{equation}
where $ \mathcal{L}_{t} $ is the Laplace transform of $ \mathcal{T}(t)
$:
\[
\mathcal{L}_{t}(\lambda)=\E\bigl( e^{-\lambda\mathcal{T}(t)}
\bigr).
\]

%
\subsection{Statistical problem}
\label{SP}
In this paper, we are going to study the problem of estimating the
characteristics of the L\'evy process $ L $ from low-frequency
observations $ Y_{0}, Y_{\Delta},\ldots,Y_{n\Delta} $ of the process $
Y $ for some fixed $ \Delta>0$. Moving to the spectral domain and
taking into account (\ref{psi}), we can reformulate our problem as the
problem of semi-parametric estimation of the characteristic exponent $
\psi$ under structural assumption (\ref{psi}) from an estimate of $
\phi_{Y}(u|\Delta) $ based on $ Y_{0}, Y_{\Delta},\ldots,Y_{n\Delta
}$.
The formula (\ref{CFY}) shows that the function $ \phi_{Y}(u|\Delta
) $
can be viewed as a composite function and our statistical problem is
hence closely related to the problem of statistical inference on the
components of a composite function. The latter type of problems in
regression setup has gotten much attention recently [see, e.g.,
\citet
{HM} and \citet{JLT}]. Our problem has, however, some features not
reflected in the previous literature. First, the unknown link function
$ \mathcal{L}_{\Delta}$, being the Laplace transform of the r.v. $
\mathcal{T}(\Delta)$, is completely monotone. Second, the
complex-valued function $ \psi$ is of the form~(\ref{psi}) implying,
for example, a certain asymptotic behavior of $ \psi(u) $ as $ u\to
\infty$. Finally, we are not in regression setup and $ \phi
_{Y}(u|\Delta) $ is to be estimated by its empirical counterpart
\[
\widehat\phi(u)=\frac{1}{n}\sum_{j=1}^{n}e^{ \ii u^{\top}(
Y_{\Delta
j}-Y_{\Delta(j-1)})}.
\]
The contribution of this paper to the literature on composite function
estimation is twofold. On the one hand, we introduce and study a new
type of statistical problems which can be called estimation of a
composite function under structural constraints. On the other hand, we
propose new and constructive estimation approach which is rather
general and can be used to solve other open statistical problems of
this type. For example, one can directly adapt our method to the
problem of semi-parametric inference in distributional Archimedian
copula-based models [see, e.g., \citet{MN} for recent results],
where one faces the problem of estimating a multidimensional
distribution function of the form
\[
F(x_{1},\ldots,x_{d})=G\bigl(f_{1}(x_{1})+\cdots+f_{d}(x_{d})\bigr),\qquad
(x_{1},\ldots,x_{d})\in\mathbb{R}^{d},
\]
with a completely monotone function $ G $ and some functions $
f_{1},\ldots, f_{d}$. Further discussion on the problem of composite
function estimation can be found in Remark \ref{compfuncbounds}.

\subsection{Specification analysis}
\label{SA}
It is clear that without further restrictions on the class of
time-changed L\'evy processes our problem of estimating $ \nu$ is not
well defined, as even in the case of the perfectly known distribution
of the process $ Y $ the parameters of the L\'evy process $ L $ are
generally not identifiable. Moreover, the corresponding statistical
procedure will suffer from the ``curse of dimensionality'' as the
dimension $ d $ increases. In order to avoid these undesirable
features, we have to impose some additional restrictions on the
structure of the time-changed process~$Y$. In statistical literature,
one can basically find two types of restricted composite models:
additive models and single-index models. While the latter class of
models is too restrictive in our situation, the former one naturally
appears if one assumes the independence of the components of $ L_{t} $.
In this paper,
we study a class of time-changed L\'evy processes satisfying the
following two assumptions:
\begin{longlist}[(ALI)]
\item[(ALI)] The L\'evy process $ L_{t} $ has independent components
such that at least two of them are nonzero, that is,
%
%
\begin{equation}
\label{CFYA}
\phi_{Y}(u|t)=\mathcal{L}_{t}\bigl(-\psi_{1}(u_{1})-\cdots-\psi_{d}(u_{d})\bigr),
\end{equation}
where $ \psi_{k}, k=1,\ldots,d$, are the characteristic exponents of
the components of~$L_{t}$ of the form
%
%
\begin{eqnarray}
\label{psik}
\psi_{k}(u)&=&\ii\mu_{k}u-\sigma_{k}^{2}u^{2}/2\nonumber\\[-8pt]\\[-8pt]
&&{} +
\int_{\mathbb{R}}\bigl( e^{\ii ux}-1-\ii ux\cdot\mathbf{1}_{\{
|x|\leq
1 \}} \bigr)\nu_{k}(dx),\qquad k=1,\ldots,d,\nonumber
\end{eqnarray}
and
%
%
\begin{equation}
\label{IA}
|\mu_{l}|+\sigma^{2}_{l}+\int_{\mathbb{R}} x^{2}\nu_{l}(dx) \neq0
\end{equation}
for at least two different indexes $ l$.
\item[(ATI)] The time change process $ \mathcal{T} $ is independent of
the L\'evy process $L$ and satisfies $ \E[\mathcal{T}(t)]=t$.
\end{longlist}

\subsubsection*{Discussion}
The advantage of the modeling
framework (\ref{CFYA}) is twofold. On the one hand, models of this
type are rather flexible: the distribution of~$Y_{t}$ for a fixed $ t
$ is in general determined by $ d+1 $ nonparametric components and $
2\times d $ parametric ones. On the other hand, these models remain
parsimonious and, as we will see later, admit statistical inference not
suffering from the ``curse of dimensionality'' as $ d $ becomes large.
The latter feature of our model is in accordance with the well
documented behavior of the additive models in regression setting and
may become particularly important if one is going to use it, for
instance, to model large portfolios of assets. The nondegeneracy
assumption (\ref{IA}) basically excludes one-dimensional models and
is\vadjust{\goodbreak}
not restrictive since it can be always checked prior to estimation by
testing that
\[
-\partial_{u_{l}u_{l}}\widehat\phi(u)|_{u=0}=\frac{1}{n}\sum
_{j=1}^{n}\bigl(Y_{\Delta j,l}-Y_{\Delta(j-1),l}\bigr)^{2}> 0
\]
for at least two different indexes $ l$.
Let us make a few remarks on the one-dimensional case, where
%
%
\begin{equation}
\label{CFY1D}
\phi_{Y}(u|t)=\mathcal{L}_{t}(-\psi_{1}(u)),\qquad t\geq0.
\end{equation}
If $ \mathcal{L}_{\Delta} $ is known, that is, the distribution of the
r.v. $ \mathcal{T}(\Delta) $ is known, we can
consistently estimate the L\'evy measure $ \nu_{1} $ by inverting $
\mathcal{L}_{\Delta} $ (see Section \ref{ext} for more details). In the
case when the function $ \mathcal{L}_{\Delta} $ is unknown, one needs
some additional assumptions (e.g., absolute continuity of the time
change) to ensure identifiability. Indeed, consider a class of the
one-dimensional L\'evy processes of the so-called compound exponential
type with the characteristic exponent of the form
\[
\psi(u)=\log\biggl[ \frac{1}{1-\widetilde\psi(u)} \biggr],
\]
where $ \widetilde\psi(u) $ is the characteristic exponent of another
one-dimensional L\'evy process $ \widetilde{L}_{t}$.
It is well known [see, e.g., Section 3 in Chapter 4 of \citet{SV}]
that $ \exp(\psi(u)) $ is the characteristic function of some
infinitely divisible distribution if $ \exp(\widetilde\psi(u)) $
does. Introduce
\[
\widetilde{\mathcal{L}}_{\Delta}(z)=\mathcal{L}_{\Delta}\bigl(\log(1+z)\bigr).
\]
As can\vspace*{1pt} be easily seen, the function $ \widetilde{\mathcal{L}}_{\Delta}
$ is completely monotone with
\mbox{$ \widetilde{\mathcal{L}}_{\Delta}(0)\,{=}\,1 $} and $ \widetilde{\mathcal
{L}}'_{\Delta}(0)= \mathcal{L}'_{\Delta}(0)$.
Moreover, it is fulfilled $ \widetilde{\mathcal{L}}_{\Delta
}(-\widetilde\psi(u))=\mathcal{L}_{\Delta}(-\psi(u)) $ for all $
u\in
\mathbb{R}$.
The existence of the time change (increasing) process $ \mathcal{T} $
with a given marginal $ \mathcal{T}(\Delta) $ can be derived from the
general theory of stochastic partial ordering [see \citet{KK}].
The above construction indicates that the assumption $ \E[\mathcal
{T}(t)]=t, t\geq0$, is not sufficient to ensure the
identifiability in the case of one-dimensional time-changed L\'evy models.

\vspace*{-3pt}\section{Estimation}\vspace*{-3pt}
\label{EST}
\subsection{Main ideas}
Assume that the L\'evy measures of the component processes $
L^{1}_{t},\ldots, L_{t}^{d} $ are absolutely continuous with integrable
densities $ \nu_{1}(x),\ldots,\allowbreak\nu_{d}(x) $ that satisfy
\[
\int_{\mathbb{R}} x^{2}\nu_{k} (x) \,dx<\infty,\qquad k=1,\ldots, d.
\]
Consider the functions
\[
\bar\nu_{k}(x)=x^{2}\nu_{k}(x),\qquad k=1,\ldots, d.
\]
By differentiating $ \psi_{k} $ two times, we get
\[
\psi''_{k}(u)=-\sigma_{k}^{2}-\int_{\mathbb{R}} e^{\ii ux} \bar\nu
_{k}(x) \,dx.
\]
For the sake of simplicity, in the sequel we will make the following
assumption:
\begin{longlist}[(ALS)]
\item[(ALS)] The diffusion volatilities $ \sigma_{k}, k=1,\ldots, d$,
of the L\'evy process $L$ are supposed to be known.
\end{longlist}
A way how to extend our results to the case of the unknown
$ (\sigma_{k}) $ is outlined in Section \ref{ext}.
Introduce the functions $ \bar\psi_{k}(u)=\psi_{k}(u)+\sigma
_{k}^{2}u^{2}/2 $ to get
%
%
\begin{equation}
\label{FourierNu}
\mathbf{F}[\bar\nu_{k}](u) =-\bar\psi''_{k}(u)=-\sigma
_{k}^{2}-\psi''_{k}(u),
\end{equation}
where $\mathbf{F}[\bar\nu_{k}](u)$ stands for the Fourier transform of
$\bar\nu_{k}$.
Denote $ Z=Y_{\Delta}$, $ \phi_{k}(u)=\partial_{u_{k}}\phi_{Z}(u),
\phi
_{kl}(u)=\partial_{u_k u_{l}}\phi_{Z}(u) $ and $\phi
_{jkl}(u)=\partial
_{u_{j}u_k u_{l}}\phi_{Z}(u) $ for $ j,k$, $l\in\{ 1,\ldots, d \}$ with
%
%
\begin{equation}
\label{PHIZ}
\phi_{Z}(u)=\E[ \exp(\ii u^{\top} Z) ]=\mathcal
{L}_{\Delta
}\bigl(-\psi_{1}(u_{1})-\cdots-\psi_{d}(u_{d})\bigr).
\end{equation}
Fix some $ k\in\{ 1,\ldots, d \} $ and for any real number $ u $
introduce a vector
\[
u^{(k)}=(0,\ldots, 0,u,0,\ldots,0)\in\mathbb{R}^{d}
\]
with $ u $ being placed at the $ k $th coordinate of the vector $
u^{(k)}$. Choose some $ l\neq k$, such that the component $ L^{l}_{t}
$ is not degenerated. Then we get from~(\ref{PHIZ})
%
%
\begin{equation}
\label{PhiRatio1}
\frac{\phi_{k}(u^{(k)})}{\phi_{l}(u^{(k)})}=\frac{\psi
'_{k}(u)}{\psi'_{l}(0)},
\end{equation}
if $ \mu_{l}\neq0 $
and
%
%
\begin{equation}
\label{PhiRatio2}
\frac{\phi_{k}(u^{(k)})}{\phi_{ll}(u^{(k)})}=\frac{\psi
'_{k}(u)}{\psi''_{l}(0)}
\end{equation}
in the case $ \mu_{l}=0$.
The identities $ \phi_{l}(\mathbf{0})=-\psi'_{l}(0)\mathcal
{L}'_{\Delta
}(0) $ and $ \phi_{ll}(\mathbf{0})=[\psi'_{ l}(0)]^{2}\times\mathcal
{L}''_{\Delta}(0)-\psi''_{l}(0)\mathcal{L}'_{\Delta}(0) $ imply
$ \psi'_{l}(0)=- [\mathcal{L}'_{\Delta}(0)]^{-1}\phi_{l}(\mathbf
{0})=\Delta^{-1}\phi_{l}(\mathbf{0})$ and
$ \psi''_{l}(0) =- [\mathcal{L}'_{\Delta}(0)]^{-1}\phi_{ll}(\mathbf
{0})=\Delta^{-1}\phi_{ll}(\mathbf{0})$ if $ \psi'_{l}(0)=0$, since
$ \mathcal{L}'_{\Delta}(0)=-\E[\mathcal{T}(\Delta)]=-\Delta$.
Combining this with (\ref{PhiRatio1}) and (\ref{PhiRatio2}), we derive
%
%
\begin{eqnarray}\hspace*{26pt}
\label{PhiDeriv21}
\psi''_{k}(u)&=&\Delta^{-1}\phi_{l}(\mathbf{0})\frac{\phi
_{kk}(u^{(k)})\phi_{l}(u^{(k)})-\phi_{k}(u^{(k)})
\phi_{lk}(u^{(k)})}{\phi^{2}_{l}(u^{(k)})},\qquad \mu_{l}\neq0,
\\
\label{PhiDeriv22}
\psi''_{k}(u)&=&\Delta^{-1}\phi_{ll}(\mathbf{0})\frac{\phi
_{kk}(u^{(k)})\phi_{ll}(u^{(k)})-\phi_{k}(u^{(k)})
\phi_{llk}(u^{(k)})}{\phi^{2}_{ll}(u^{(k)})},\qquad \mu_{l}= 0.
\end{eqnarray}
Note that in the above derivations we have repeatedly used
assumption (ATI), that turns out to be crucial for the identifiability.
The basic idea of the algorithm, we shall develop in the Section \ref
{ALG}, is to estimate $ \bar\nu_{k} $
by an\vspace*{1pt} application of the regularized Fourier inversion formula to an
estimate of $ \bar\psi''_{k}(u)$.
As indicated by formulas (\ref{PhiDeriv21}) and (\ref{PhiDeriv22}), one
could, for example, estimate $ \bar\psi''_{k}(u)$, if some estimates
for the functions $ \phi_{k}(u), \phi_{lk}(u) $ and $ \phi_{llk}(u) $
are available.
\begin{rem}
One important issue we would like to comment on is the robustness of
the characterizations (\ref{PhiDeriv21}) and (\ref{PhiDeriv22}) with
respect to the independence assumption for the components of the L\'evy
process $ L_{t}$. First, note that if the components are dependent,
then the key identity (\ref{FourierNu}) is not any longer valid for $
\psi''_{k} $ defined as in (\ref{PhiDeriv21}) or (\ref{PhiDeriv22}).
Let us determine how strong can it be violated. For concreteness,
assume that $ \mu_{l}>0 $ and that the dependence in the components of
$ L_{t} $ is due to a correlation between diffusion components. In
particular, let $ \Sigma(k,l)>0$.
Since in the general case
\[
\partial_{u_{k}} \psi\bigl(u^{(k)}\bigr)=\partial_{u_{l}}\psi\bigl(u^{(k)}\bigr)\frac
{\phi_{k}(u^{(k)})}{\phi_{l}(u^{(k)})}
\]
and $ \partial_{u_{k}u_{k}} \psi(u^{(k)})=-\sigma^{2}_{k}-\mathbf
{F}[\bar\nu_{k}](u)$,
we get
\[
\mathbf{F}[\bar\nu_{k}](u)+\psi''_{k}(u)+\sigma^{2}_{k} =\frac
{\Sigma
(k,l)}{2}\biggl[ u \partial_{u_{k}}\biggl\{ \frac{\phi
_{k}(u^{(k)})}{\phi
_{l}(u^{(k)})} \biggr\}+\frac{\phi_{k}(u^{(k)})}{\phi_{l}(u^{(k)})}
\biggr].
\]
Using the fact that both functions $ u \partial_{u_{k}}\{ \phi
_{k}(u^{(k)})/\phi_{l}(u^{(k)})\} $ and
$ \phi_{k}(u^{(k)})/\allowbreak\phi_{l}(u^{(k)}) $ are uniformly bounded for $
u\in
\mathbb{R}$, we get that
the model ``misspecification bias'' is bounded by $ C\Sigma(k,l) $ with
some constant $ C>0$. Thus, the weaker is the dependence between
components $ L^{k} $ and $ L^{l} $, the
smaller is the resulting ``misspecification bias.''
\end{rem}

\subsection{Algorithm}
\label{ALG}
Set $ Z_{j}=Y_{\Delta j}-Y_{\Delta(j-1)}, j=1,\ldots, n$, and denote
by $Z_{j}^{k}$ the $k$th coordinate of $Z_{j}$. Note that $Z_{j},
j=1,\ldots, n$, are identically distributed. The estimation procedure
consists basically of three steps:
\begin{longlist}[Step 1.]
\item[Step 1.] First, we are interested in estimating partial
derivatives of the function $ \phi_{Z}(u) $ up to the third order.
To this end, define
%
%
\begin{eqnarray}
\label{PhiDeriv1Est}
\widehat\phi_{k}(u)&=&\frac{1}{n}\sum_{j=1}^{n}Z^{k}_{j}\exp(\ii
u^{\top
} Z_{j}),
\\
\label{PhiDeriv2Est}
\widehat\phi_{lk}(u)&=&\frac{1}{n}\sum
_{j=1}^{n}Z^{k}_{j}Z^{l}_{j}\exp
(\ii u^{\top} Z_{j}),
\\
\label{PhiDeriv3Est}
\widehat\phi_{llk}(u)&=&\frac{1}{n}\sum
_{j=1}^{n}Z^{k}_{j}Z^{l}_{j}Z^{l}_{j}\exp(\ii u^{\top} Z_{j}).
\end{eqnarray}
\item[Step 2.] In a second step, we estimate the second derivative of
the characteristic exponent $ \psi_{k}(u)$. Set
%
%
\begin{equation}
\label{PsiDeriv2Est1}
\widehat\psi_{k,2}(u)=\Delta^{-1}\widehat\phi_{l}(\mathbf
{0})\frac
{\widehat\phi_{kk}(u^{(k)})
\widehat\phi_{l}(u^{(k)})-\widehat\phi_{k}(u^{(k)})\widehat\phi
_{lk}(u^{(k)})}
{[\widehat\phi_{l}(u^{(k)})]^{2}},
\end{equation}
if $|\widehat\phi_{l}(\mathbf{0})|>\kappa/\sqrt{n}$ and
%
%
\begin{equation}
\label{PsiDeriv2Est2}
\widehat\psi_{k,2}(u)=\Delta^{-1}\widehat\phi_{ll}(\mathbf
{0})\frac
{\widehat\phi_{kk}(u^{(k)})
\widehat\phi_{ll}(u^{(k)})-\widehat\phi_{k}(u^{(k)})\widehat\phi
_{llk}(u^{(k)})}
{[\widehat\phi_{ll}(u^{(k)})]^{2}}
\end{equation}
otherwise,
where $ \kappa$ is a positive number.
\item[Step 3.] Finally,\vspace*{1pt} we construct an estimate for $ \bar\nu_{k}(x)
$ by applying the Fourier inversion formula
combined with a regularization to $ \widehat\psi_{k,2}(u) $:
%
%
\begin{equation}
\label{NUEST}
\widehat{\nu}_{k}(x)=-\frac{1}{2\pi}\int_{\mathbb{R}}e^{-\ii
ux}[\widehat{\psi}_{k,2}(u)+\sigma_{k}^{2}] \mathcal{K}(uh_{n})
\,du,
\end{equation}
where $ \mathcal{K}(u) $ is a regularizing kernel supported on $ [-1,1]
$ and $ h_{n} $ is a sequence of bandwidths which tends to $ 0 $ as $
n\to\infty$.
The choice of the sequence $ h_{n} $ will be discussed later on.
\end{longlist}
\begin{rem}
The parameter $ \kappa$ determines the testing error for the
hypothesis $ H\dvtx\mu_{l}>0$. Indeed, if
$ \mu_{l}=0$, then $ \phi_{l}(\mathbf{0})=0 $ and by the central
limit theorem
\begin{eqnarray*}
\mathbb{P}\bigl(|\widehat\phi_{l}(\mathbf{0})|>\kappa/\sqrt{n}
\bigr)&\leq&\mathbb{P}
\bigl(\sqrt{n}|\widehat\phi_{l}(\mathbf{0})-\phi_{l}(\mathbf
{0})|>\kappa
\bigr)\\
&\to&\mathbb{P}\bigl(|\xi|>\kappa/\sqrt{\Var[Z^{l}]}\bigr),\qquad
n\to
\infty,
\end{eqnarray*}
with $ \xi\sim\mathcal{N}(0,1)$.
\end{rem}

%
\section{Asymptotic analysis}
\label{ASYMP}
In this section, we are going to study the asymptotic properties of
the estimates $ \widehat{\nu}_{k}(x)$,
$ k=1,\ldots,d$.
In particular, we prove almost sure uniform as well as pointwise
convergence rates for $ \widehat{\nu}_{k}(x)$. Moreover, we will
show the optimality of the above rates over suitable classes of
time-changed L\'evy models.

\subsection{\texorpdfstring{Global vs. local smoothness of L\'evy
densities}{Global vs. local smoothness of Levy densities}}
\label{CLASSLEVY}
Let $ L_{t} $ be a one-dimen\-sional L\'evy process with a L\'evy
density $ \nu$. Denote $ \bar\nu(x)=x^{2}\nu(x)$.
For any two nonnegative numbers $ \beta$ and $ \gamma$ such that
$ \gamma\in[0,2] $ consider two following classes of L\'evy densities
$ \nu$:
%
%
\begin{equation}
\label{ClCond1}
\mathfrak{S}_{\beta}=\biggl\{ \nu\dvtx\int_{\mathbb{R}} (1+|u|^{\beta
})\mathbf{F}[\bar\nu](u) \,du<\infty\biggr\}
\end{equation}
and
%
%
\begin{equation}
\label{ClCond2}
\mathfrak{B}_{\gamma}=\biggl\{\nu\dvtx\int_{|y|>\epsilon}\nu(y)
\,dy\asymp
\frac{\Pi(\epsilon)}{\epsilon^{\gamma}}, \epsilon\to+0
\biggr\},
\end{equation}
where $ \Pi$ is some positive function on $ \mathbb{R}_{+} $
satisfying $ 0<\Pi(+0)<\infty$.
The parameter $ \gamma$
is usually called the Blumenthal--Geetor
index of $ L_{t} $.
This index~$\gamma$ is related to the ``degree of activity'' of jumps
of $ L_{t} $. All L\'evy measures put finite mass
on the set $(-\infty,-\epsilon]\cup[\epsilon,\infty) $ for any
arbitrary $\epsilon>0$.
If $\nu([-\epsilon,\epsilon])<\infty$
the process
has finite activity and $\gamma=0$.
If $ \nu([-\epsilon,\epsilon])=\infty$,
that is, the process
has infinite activity and in addition the L\'evy measure
$\nu((-\infty,-\epsilon]\cup[\epsilon,\infty))$ diverges near $0$
at a rate $|\epsilon|^{-\gamma} $ for some $\gamma>0$, then the
Blumenthal--Geetor
index of $ L_{t} $ is equal to $\gamma$. The higher $\gamma$ gets, the
more frequent the
small jumps become.

Let us now investigate the connection between classes $ \mathfrak
{S}_{\beta} $ and $ \mathfrak{B}_{\gamma} $. First, consider an
example. Let $ L_{t} $ be a tempered stable L\'evy process
with a L\'evy density
\[
\nu(x)=\frac{2^{\gamma}\cdot\gamma}{\Gamma(1-\gamma)}x^{-(\gamma
+1)}\exp\biggl( -\frac{x}{2} \biggr)\mathbf{1}_{(0,\infty)}(x),\qquad
x>0,
\]
where $ \gamma\in(0,1)$. It is clear that $ \nu\in\mathfrak
{B}_{\gamma}$ but what is about $ \mathfrak{S}_{\beta} $? Since
\[
\bar\nu(x)=\frac{2^{\gamma}\cdot\gamma}{\Gamma(1-\gamma
)}x^{1-\gamma
}\exp\biggl( -\frac{x}{2} \biggr)\mathbf{1}_{(0,\infty)}(x),
\]
we derive
\[
\mathbf{F}[\bar\nu](u)=\int_{0}^{\infty} e^{\ii u x}\bar\nu(x)
\,dx\asymp2^{\gamma}\gamma(1-\gamma)e^{\ii\pi(1-\gamma/2)
}u^{-2+\gamma
},\qquad u\to+\infty,
\]
by the Erd\'elyi lemma [see \citet{ER}]. Hence, $ \nu$ cannot belong
to $\mathfrak{S}_{\beta} $ as long as $ \beta>1-\gamma$. The message
of this example is that given the activity index $ \gamma$, the
parameter $ \beta$ determining the smoothness of
$ \bar\nu$, cannot be taken arbitrary large. The above example can
be straightforwardly generalized to a class of L\'evy densities
supported on $ \mathbb{R}_{+}$. It turns out that if the L\'evy
density $ \nu$ is supported on $ [0,\infty) $, is infinitely smooth
in $ (0,\infty) $ and $ \nu\in\mathfrak{B}_{\gamma}$ for some $
\gamma\in(0,1)$, then $ \nu\in\mathfrak{S}_{\beta} $ for all $
\beta$ satisfying $ 0\leq\beta<1-\gamma$ and
$ \nu\notin\mathfrak{S}_{\beta}$ for $ \beta>1-\gamma$. As a
matter of fact, in the case $ \gamma=0 $ (finite activity case) the
situation is different and $ \beta$ can be arbitrary large.

The above discussion indicates that in the case $ \nu\in\mathfrak
{B}_{\gamma}$ with some $ \gamma>0 $ it is reasonable to look at the
local smoothness
of $ \bar\nu_{k} $ instead of the global one. To this end, fix a point
$ x_{0}\in\mathbb{R} $ and a positive integer number $ s\geq1$. For
any $\delta>0 $ and $D>0$ introduce a class $ \mathfrak
{H}_{s}(x_{0},\delta,D) $ of L\'evy densities $ \nu$ defined as
%
%
\begin{eqnarray}
\label{pnormx0}
\mathfrak{H}_{s}(x_{0},\delta,D)
&=&\Bigl\{\nu\dvtx\bar\nu(x)\in
C^{s}(]x_{0}-\delta,x_{0}+\delta[),\nonumber\\[-8pt]\\[-8pt]
&&\hphantom{\Bigl\{}
\sup
_{x\in]x_ {0}-\delta,x_{0}+\delta[}\bigl|\bar\nu^{(l)}(x)\bigr| \leq D \mbox{
for } l=1,\ldots, s\Bigr\}.
\nonumber
\end{eqnarray}

%
\subsection{Assumptions}
\label{ASS}
In order to prove the convergence of $ \widehat{\nu}_{k}(x)$, we need
the assumptions listed below:
\begin{longlist}[(AL1)]
\item[(AL1)] The L\'evy densities $ \nu_{1},\ldots, \nu_{d} $ are in
the class $ \mathfrak{B}_{\gamma} $
for some $ \gamma>0$.
\item[(AL2)] For some $ p>2$, the L\'evy densities $ \nu_{k},
k=1,\ldots,d$, have finite absolute moments of the order $ p$:
\[
\int_{\mathbb{R}} |x|^{p}\nu_{k}(x) \,dx<\infty,\qquad k=1,\ldots,d.
\]
\item[(AT1)] The time change $\mathcal{T}$ is independent of the L\'evy
process $L$ and the sequence $ T_{k}=\mathcal{T}(\Delta k)-\mathcal
{T}(\Delta(k-1)),
k\in\mathbb{N}$, is strictly stationary, $ \alpha$-mixing with the
mixing coefficients $ (\alpha_{T}(j))_{j\in\mathbb{N}} $ satisfying
\[
\alpha_{T}(j)\leq\bar\alpha_{0}\exp(-\bar\alpha_{1} j),\qquad
j\in
\mathbb{N},
\]
for some positive constants $ \bar\alpha_{0} $ and $ \bar\alpha
_{1}$.
Moreover, assume that
\[
\E[\mathcal{T}^{-2/\gamma}(\Delta)]<\infty,\qquad
\E[\mathcal{T}^{2p}(\Delta)]<\infty
\]
with $ \gamma$ and $ p $ being from  assumptions (AL1) and (AL2),
respectively.
\item[(AT2)] The Laplace transform $ \mathcal{L}_{t}(z) $ of $
\mathcal
{T}(t) $ fulfills
\[
\mathcal{L}'_{t}(z)=o(1),\qquad \mathcal{L}''_{t}(z)/\mathcal
{L}'_{t}(z)=O(1),\qquad |z| \to\infty,\qquad \Ree z>0.
\]
\item[(AK)] The regularizing kernel $ \mathcal{K} $ is uniformly
bounded, is supported on $ [-1,1] $ and satisfies
\[
\mathcal{K}(u)=1,\qquad u\in[-a_{K},a_{K}],
\]
with some $ 0<a_{K}<1$.
\item[(AH)] The sequence of bandwidths $ h_{n} $ is assumed to satisfy
\[
h^{-1}_{n}=O(n^{1-\delta}),\qquad M_{n}\sqrt{\frac{\log n }{n}}\sqrt
{\frac{1}{h_{n}}\log\frac{1}{h_{n}}}=o(1),\qquad n\to\infty,
\]
for some positive number $ \delta$ fulfilling $ 2/p<\delta\leq1$, where
\[
M_{n}= \max_{l\neq
k}\sup_{\{|u|\leq1/h_{n}\}}\bigl|\phi^{-1}_{l}\bigl(u^{(k)}\bigr)\bigr|.
\]
\end{longlist}
\begin{rem}
By requiring $ \nu_{k}\in\mathfrak{B}_{\gamma}, k=1,\ldots, d$,
with some $ \gamma>0$, we exclude from our analysis pure compound
Poisson processes and some infinite activity L\'evy processes with $
\gamma=0$. This is mainly done for the sake of brevity: we would like
to avoid additional technical calculations related to the fact that the
distribution of $ Y_{t} $ is not in general absolutely continuous in
this case.
\end{rem}
\begin{rem} Assumption (AT1) is satisfied if, for example, the process~$ \mathcal{T}(t) $ is of the form (\ref{TCD}),
where the rate process $ \rho(u) $ is strictly\vadjust{\goodbreak} stationary,
geometrically $ \alpha$-mixing and fulfills
%
%
\begin{equation}
\label{momassinteg}
\E[\rho^{2p}(u)]<\infty,\qquad u\in[0,\Delta],\qquad \E\biggl(
\int
_{0}^{\Delta}\rho(u) \,du \biggr)^{-2/\gamma}<\infty.
\end{equation}
In the case of the Cox--Ingersoll--Ross process $ \rho$ (see Section
\ref{sec52}),  assumptions (\ref{momassinteg}) are satisfied for any
$ p>0
$ and any $ \gamma>0$.
\end{rem}
\begin{rem}
Let us comment on  assumption (AH). Note that in order to determine
$M_{n}$, we do not need the characteristic function $ \phi(u) $ itself,
but only a low bound for its tails. Such low bound can be constructed
if, for example, a low bound for the tail of $ \mathcal{L}'_{t}(z) $
and an upper bound for the Blumenthal--Geetor index $ \gamma$ are
available [see \citet{Bel1} for further discussion]. In practice,
of course, one should prefer adaptive methods for choosing $ h_{n}$.
One such method, based on the so called ``quasi-optimality'' approach,
is proposed and used in Section \ref{TCGamma}. The theoretical analysis
of this
method is left for  future research.
\end{rem}

%
\subsection{Uniform rates of convergence}
\label{URC}
Fix some $ k $ from the set $ \{ 1,2,\ldots,d \}$. Define a weighting function
$ w(x)=\log^{-1/2}(e+|x|) $ and denote
\[
\|\bar\nu_{k}-\widehat\nu_{k}\|_{L_{\infty}(\mathbb
{R},w)}=\sup_{x\in\mathbb{R}}[w(|x|)|\bar\nu_{k}(x)-\widehat\nu_{k}(x)|].
\]
Let $ \xi_{n} $ be a sequence of positive r.v. and $ q_{n} $ be a
sequence of positive real numbers.
We shall write $ \xi_{n}=O_{\mathrm{a.s.}}(q_{n}) $ if there is a constant $ D>0
$ such that $ \mathbb{P}(\limsup_{n\to\infty} q_{n}^{-1}\xi_{n}\leq
D)=1$. In
the case $ \mathbb{P}(\limsup_{n\to\infty} q_{n}^{-1}\xi_{n}=0)=1 $, we shall
write $ \xi_{n}=o_{\mathrm{a.s}.}(q_{n})$.
\begin{theorem}
\label{UpperBounds}
Suppose that  assumptions \textup{(AL1)}, \textup{(AL2), (AT1), (AT2)}, \textup{(AK)}
and \textup{(AH)}
are fulfilled.
Let $ \widehat{\nu}_{k}(x) $ be the estimate for $\bar\nu_{k}(x) $
defined in Section~\ref{ALG}.
If $ \nu_{k}\in\mathfrak{S}_{\beta} $
for some $ \beta>0$, then
\[
\|\bar\nu_{k}-\widehat\nu_{k}\|_{L_{\infty}(\mathbb
{R},w)}=O_{\mathrm{a.s}.}\Biggl( \sqrt{\frac{\log^{3+\varepsilon} n}{n}\int
_{-1/h_{n}}^{1/h_{n}}\mathfrak{R}^{2}_{k}(u) \,du}+ h_{n}^{\beta
}\Biggr)
\]
for arbitrary small $ \varepsilon>0$, where
\[
\mathfrak{R}_{k}(u)=\frac{(1+|\psi'_{k}(u)|)^{2}}{|\mathcal
{L}'_{\Delta
}(-\psi_{k}(u))|}.
\]
\end{theorem}
\begin{cor}
\label{UPPERCOR1}\label{UPPERCOR2}
Suppose that $ \sigma_{k}=0$, $ \gamma\in(0,1] $ in  assumption
\textup{(AL1)}
and
\[
|\mathcal{L}'_{\Delta}(z)|\gtrsim\exp(-a|z|^{\eta}),\qquad |z|\to
\infty,\qquad \Ree z\geq0,
\]
for some $ a>0 $ and $ \eta>0$. If $ \mu_{k}>0$, then
%
%
\begin{equation}
\label{UBE1}
\|\bar\nu_{k}-\widehat\nu_{k}\|_{L_{\infty}(\mathbb
{R},w)}=O_{\mathrm{a.s}.}\Biggl(\sqrt{\frac{\log^{3+\varepsilon} n}{n}}\exp
(ac\cdot h_{n}^{-\eta} ) +h_{n}^{\beta} \Biggr)
\end{equation}
with some constant $ c>0$.
In the case $ \mu_{k}=0 $ we have
%
%
\begin{equation}
\label{UBE2}
\|\bar\nu_{k}-\widehat\nu_{k}\|_{L_{\infty}(\mathbb
{R},w)}=O_{\mathrm{a.s}.}\Biggl(\sqrt{\frac{\log^{3+\varepsilon} n}{n}}\exp
(ac\cdot h_{n}^{-\gamma\eta} ) +h_{n}^{\beta} \Biggr).
\end{equation}
Choosing $ h_{n} $ in such a way that the r.h.s. of (\ref
{UBE1}) and (\ref{UBE2}) are minimized, we obtain
the rates shown in the Table \ref{UCR}.
%
%
%
\begin{table}
\tabcolsep=0pt
\caption{Uniform convergence rates for $ \widehat\nu_{k} $ in the case
$ \sigma_{k}=0 $}\label{UCR}
\begin{tabular*}{\tablewidth}{@{\extracolsep{4in minus 4in}}llcc@{}}
\hline
\multicolumn{2}{@{}c}{$\bolds{
|\mathcal{L}'_{\Delta}(z)|\bm{\gtrsim}
|z|^{-\alpha}}
$} & \multicolumn{2}{c@{}}{$\bolds{ |\mathcal{L}'_{\Delta}(z)|\bm{\gtrsim}\exp
(-a|z|^{\eta}) }$}\\[-4pt]
\multicolumn{2}{@{}c}{\hrulefill} & \multicolumn{2}{c@{}}{\hrulefill} \\
$\bolds{ \mu_{k}>0} $ & \multicolumn{1}{c}{$ \bolds{\mu_{k}=0} $} & $ \bolds{\mu_{k}>0} $
& $ \bolds{\mu_{k}=0} $ \\
\hline
$ n^{-{\beta}/{(2\alpha+2\beta+1)}}$ & $ n^{-{\beta}/{(2\alpha
\gamma+2\beta+1)}} $ & $ \log^{-\beta/\eta} n $ & $ \log^{-\beta
/\gamma
\eta} n $\\
$\quad{}\times\log^{{(3+\varepsilon
)\beta}/{(2\alpha+2\beta+1)}}(n) $ & $\quad{}\times\log^{{(3+\varepsilon)\beta}/{(2\alpha
\gamma
+2\beta+1)}}(n)$\\
\hline
\end{tabular*}
\end{table}
%
%
%
\begin{table}[b]
\tablewidth=270pt
\caption{Uniform convergence rates for $ \widehat\nu_{k} $ in the case
$ \sigma_{k}>0 $} \label{RCS}
\begin{tabular*}{\tablewidth}{@{\extracolsep{4in minus 4in}}cc@{}}
\hline
$ \bolds{|\mathcal{L}'_{\Delta}(z)|\bm{\gtrsim}|z|^{-\alpha}} $ & $ \bolds{|\mathcal
{L}'_{\Delta}(z)|\bm{\gtrsim}\exp(-a|z|^{\eta}) }$\\
\hline
$ n^{-{\beta}/{(4\alpha+2\beta+1)}}\log^{{(3+\varepsilon
)\beta}/{(4\alpha+2\beta+1)}}(n) $ & $ \log^{-\beta/2\eta} n $ \\
\hline
\end{tabular*}
\end{table}
If $ \gamma\in(0,1] $ in  assumption \textup{(AL1)} and
\[
|\mathcal{L}'_{\Delta}(z)|\gtrsim|z|^{-\alpha},\qquad
|z|\to\infty,\qquad
\Ree z\geq0,
\]
for some $ \alpha>0$, then
\[
\|\bar\nu_{k}-\widehat\nu_{k}\|_{L_\infty(\mathbb{R},w)}
=O_{\mathrm{a.s}.}
\Biggl(\sqrt{\frac{\log^{3+\varepsilon} n}{n}} h_{n}^{-1/2-\alpha}
+h_{n}^{\beta} \Biggr)
\]
provided $ \mu_{k}>0$. In the case $ \mu_{k}=0 $, one has
\[
\|\bar\nu_{k}-\widehat\nu_{k}\|_{L_\infty(\mathbb{R},w)}
=O_{\mathrm{a.s}.}
\Biggl(\sqrt{\frac{\log^{3+\varepsilon} n}{n}} h_{n}^{-1/2-\alpha\gamma}
+h_{n}^{\beta} \Biggr).
\]
The choices $ h_{n}=n^{-1/(2(\alpha+\beta) +1)}\log^{(3+\varepsilon
)/(2(\alpha+\beta) +1)}(n) $
and
\[
h_{n}=n^{-1/(2(\alpha\gamma+\beta) +1)}
\log^{(3+\varepsilon)/(2(\alpha\gamma+\beta) +1)}(n)
\]
for the cases $ \mu_{k}>0 $ and $ \mu_{k}=0$, respectively, lead to
the bounds shown in Table~\ref{UCR}. In the case $ \sigma_{k}>0 $, the
rates of convergence are given in Table \ref{RCS}.\vadjust{\goodbreak}
\end{cor}

\begin{rem}
As one can see,  assumption (AH) is always fulfilled for the optimal
choices of $ h_{n} $ given in Corollary \ref{UPPERCOR2}, provided $
\alpha\gamma+\beta>0 $
and $ p>2+1/(\alpha\gamma+\beta)$.
\end{rem}

%
\subsection{Pointwise rates of convergence}
\label{PCR}
Since the transformed L\'evy density~$ \bar\nu_{k} $ is usually not
smooth at $ 0 $ (see Section \ref{CLASSLEVY}), pointwise rates of
convergence might be more informative than the uniform ones if $ \nu
_{k}\in\mathfrak{B}_{\gamma} $ for some $ \gamma>0$. It is remarkable
that the same estimate $ \widehat\nu_{k} $ as before will achieve the
optimal pointwise convergence rates in the class $ \mathfrak
{H}_{s}(x_{0},\delta,D)$, provided the kernel $ \mathcal{K} $
satisfies (AK) and is sufficiently smooth.
\begin{theorem}
\label{pointwiseupper}
Suppose that  assumptions \textup{(AL1), (AL2), (AT1), (AT2), (AK)} and
\textup{(AH)}
are fulfilled.
If $ \nu_{k}\in\mathfrak{H}_{s}(x_{0},\delta,D) $ with $\mathfrak
{H}_{s}(x_{0},\delta,D)$ being defined in (\ref{pnormx0}),
for some $ s\geq1, \delta>0, D>0$, and $ \mathcal{K}\in
C^{m}(\mathbb{R}) $ for some $ m\geq s$, then
%
%
\begin{equation}
\label{upperineq}
|\widehat\nu_{k}(x_{0})-\bar\nu_{k}(x_{0})|= O_{\mathrm{a.s}.}\Biggl( \sqrt
{\frac
{\log^{3+\varepsilon} n}{n}\int_{-1/h_{n}}^{1/h_{n}}\mathfrak
{R}^{2}_{k}(u) \,du}+ h_{n}^{s}\Biggr)
\end{equation}
with $ \mathfrak{R}_{k}(u) $ as in Theorem \ref{UpperBounds}.
As a result, the pointwise rates of convergence for different
asymptotic behaviors of the Laplace transform $ \mathcal{L}_{t} $
coincide with ones given in Tables \ref{UCR} and \ref{RCS}, if we
replace $ \beta$ with $ s$.
\end{theorem}
\begin{rem}
If the kernel $ \mathcal{K} $ is infinitely smooth, then it will
automatically ``adapt'' to the pointwise smoothness of $\bar\nu_{k}$,
that is, (\ref{upperineq}) will hold for arbitrary large $ s\geq1$,
provided $ \nu_{k}\in\mathfrak{H}_{s}(x_{0},\delta,D) $ with some
$\delta>0$ and $D>0$. An example of infinitely smooth kernels
satisfying (AK) is given by the so called flat-top kernels (see
Section \ref{TCGamma} for the definition).
\end{rem}

%
\subsection{Lower bounds}
\label{LBOUNDS}
In this section, we derive a lower bound on the minimax risk of an
estimate $ \widehat\nu(x) $
over a class of one-dimensional time-changed L\'evy processes $
Y_{t}=L_{\mathcal{T}(t)} $ with the known distribution
of $ \mathcal{T}$,
such that the L\'evy measure $ \nu$ of the L\'evy process $ L_{t} $
belongs to the class $ \mathfrak{S}_{\beta}\cap\mathfrak{B}_{\gamma}$
with some $ \beta>0 $ and $ \gamma\in(0,1]$.
The following theorem holds.

\begin{theorem}
\label{LowBounds}
Let $ L_{t} $ be a L\'evy process with zero diffusion part,
a~drift~$\mu$ and a L\'evy density $ \nu$. Consider a time-changed L\'evy
process $ Y_{t}=L_{\mathcal{T}(t)}$, where the Laplace transform of
the time change $ \mathcal{T}(t) $ fulfills
%
%
\begin{equation}
\label{LCond}
\mathcal{L}_{\Delta}^{(k+1)}(z)/\mathcal{L}_{\Delta}^{(k)}(z)
=O(1),\qquad |z|\rightarrow\infty,\qquad \Ree z\geq0,
\end{equation}
for $ k=0,1,2$, and uniformly in $ \Delta\in[0,1]$. Then
%
%
\begin{equation}
\quad
\label{LB}
\liminf_{n\to\infty}\inf_{\widehat\nu} \sup_{\nu\in\mathfrak
{S}_{\beta}\cap\mathfrak{B}_{\gamma} }\mathbb{P}_{(\nu,\mathcal
{T})}\bigl( \|
\bar\nu-\widehat\nu\|_{L_\infty(\mathbb{R},w)}> \varepsilon
h^{\beta
}_{n}\log^{-1}(1/h_{n})\bigr)>0
\end{equation}
for any $ \varepsilon>0 $ and any sequence $ h_{n} $ satisfying
\[
n\Delta^{-1}[ \mathcal{L}_{\Delta}^{\prime}(c\cdot
h_{n}^{-\gamma
})%
] ^{2}h_{n}^{2\beta+1 }=O(1),\qquad n\to\infty,
\]
in the case $ \mu=0 $
and
\[
n\Delta^{-1}[ \mathcal{L}_{\Delta}^{\prime}(c\cdot h_{n}^{-1
})%
] ^{2}h_{n}^{2\beta+1 }=O(1),\qquad n\to\infty,
\]
in the case $ \mu>0$, with some positive constant $ c>0$.
Note that the infimum in (\ref{LB}) is taken over all estimators of $
\nu$ based on $ n $ observations of the r.v.~$Y_{\Delta} $ and $ \mathbb{P}
_{(\nu,\mathcal{T})} $ stands for the distribution of $ n $ copies of $
Y_{\Delta}$.
\end{theorem}
\begin{cor}
\label{ExpL}
Suppose that the underlying L\'evy process is driftless, that is, $ \mu=0
$ and $ \mathcal{L}_{t}(z)=\exp(-azt) $ for some $ a>0$,
corresponding to a~deterministic time change process $ \mathcal{T}(t)=at$. Then by taking
\[
h_{n}=\biggl( \frac{\log n - ((2\beta+1)/\gamma) \log\log
n}{2ac\Delta
} \biggr)^{-1/\gamma},
\]
we arrive at
\[
\liminf_{n\to\infty}\inf_{\widehat\nu} \sup_{\nu\in\mathfrak
{S}_{\beta}\cap\mathfrak{B}_{\gamma} }\mathbb{P}_{(\nu,\mathcal
{T})}\bigl(\|
\bar\nu-\widehat\nu\|_{L_{\infty}(\mathbb{R},w)}> \varepsilon
\cdot
\Delta^{\beta/\gamma} \log^{-\beta/\gamma} n\bigr)>0.\vspace*{-3pt}
\]
\end{cor}
\begin{cor}
\label{PolL}
Again let $ \mu=0$. Take $ \mathcal{L}_{t}(z)=1/(1+z)^{\alpha_{0}
t}, \Ree z>0 $ for some $ \alpha_{0} >0$, resulting in a Gamma
process $ \mathcal{T}(t) $ (see Section \ref{TCGamma} for the
definition). Under the choice
\[
h_{n}=(n\Delta)^{-1/(2\alpha\gamma+2\beta+1)}
\]
we get
\[
\liminf_{n\to\infty}\inf_{\widehat\nu} \sup_{\nu\in\mathfrak
{S}_{\beta}\cap\mathfrak{B}_{\gamma} }\mathbb{P}_{(\nu,\mathcal
{T})}\bigl(\|
\bar\nu-\widehat\nu\|_{L_{\infty,w}(\mathbb{R})}> \varepsilon
\cdot
(n\Delta)^{-\beta/(2\alpha\gamma+2\beta+1)}\log^{-1}n\bigr)>0,\vspace*{-3pt}
\]
where $ \alpha= \alpha_{0} \Delta+1$.
\end{cor}
\begin{rem}
\label{HighFreq}
 Theorem \ref{LowBounds} continues to hold for $ \Delta\to0 $ and
therefore can be used to derive minimax lower bounds for the risk of $
\widehat\nu$ in high-frequency setup. As can be seen from
Corollaries \ref{ExpL} and \ref{PolL}, the rates will strongly depend
on the specification of the time change process $
\mathcal{T}$.\vspace*{-3pt}
\end{rem}

The pointwise rates of convergence obtained in (\ref{pointwiseupper})
turn out to be optimal over the class
$ \mathfrak{H}_{s}(x_{0},\delta,D)\cap\mathfrak{B}_{\gamma} $ with $
s\geq1$,
$\delta>0$, $ x_{0}\in\mathbb{R}$, $D>0$ and $ \gamma\in(0,1] $ as
the next theorem shows.
\begin{theorem}
\label{LowBoundsPW}
$\!\!\!$Let $ L_{t} $ be a L\'evy process with zero diffusion part, a~drift~$
\mu$ and a L\'evy density $ \nu$. Consider a time-changed
L\'evy
process $ Y_{t}=L_{\mathcal{T}(t)}$, where the Laplace transform of
the time change $ \mathcal{T}(t) $ fulfills (\ref{LCond}). Then
%
%
\begin{equation}
\label{LBP}
\liminf_{n\to\infty}\inf_{\widehat\nu}\!\sup_{\nu\in\mathfrak
{H}_{s}(x_{0},\delta,D)\cap\mathfrak{B}_{\gamma} }\!\mathbb{P}_{(\nu,\mathcal
{T})}\bigl( |\bar\nu(x_{0})\,{-}\,\widehat\nu(x_{0})|\,{>}\, \varepsilon
h^{s}_{n}\log^{-1}(1/h_{n})\bigr)>0\hspace*{-37pt}\vadjust{\goodbreak}
\end{equation}
for $ s\geq1$, $\delta>0$, $D>0$, any $ \varepsilon>0 $ and any
sequence $ h_{n} $ satisfying
\[
n\Delta^{-1}[ \mathcal{L}_{\Delta}^{\prime}(c\cdot
h_{n}^{-\gamma
})%
] ^{2}h_{n}^{2s+1 }=O(1),\qquad n\to\infty,
\]
in the case $ \mu=0 $
and
\[
n\Delta^{-1}[ \mathcal{L}_{\Delta}^{\prime}(c\cdot h_{n}^{-1
})%
] ^{2}h_{n}^{2s+1 }=O(1),\qquad n\to\infty,
\]
in the case $ \mu>0$, with some positive constant $ c>0$.
\end{theorem}

%
\subsection{Extensions}
\label{ext}
\subsubsection*{\texorpdfstring{One-dimensional time-changed L\'evy models}{One-dimensional
time-changed Levy models}} Let us consider
a class of one-dimensional time-changed L\'evy models (\ref{CFY1D})
with the known time change process, that is, the known function $
\mathcal{L}_{t} $ for all $ t>0$.
This class of models trivially includes L\'evy processes without time
change [by setting $ \mathcal{L}_{t}(z)=\exp(-tz) $] studied in
\citet{NR} and \citet{CG}.
We have in this case
%
%
\begin{equation}\quad
\label{PhiDeriv23}
\psi''_{1}(u)=-\frac{\phi''(u)\mathcal{L}'_{\Delta}(-\psi
_{1}(u))-\phi
'(u)\mathcal{L}''_{\Delta}(-\psi_{1}(u))/\mathcal{L}'_{\Delta
}(-\psi
_{1}(u))}{[\mathcal{L}'_{\Delta}(-\psi_{1}(u))]^{2}}
\end{equation}
with
\[
\psi_{1}(u)=-\mathcal{L}_{\Delta}^{-}(\phi(u)),
\]
where $ \mathcal{L}_{\Delta}^{-} $ is an inverse function for $
\mathcal
{L}_{\Delta}$.
Thus, $ \psi''_{1}(u) $ is again a ratio-type estimate involving the
derivatives of the c.f. $\phi$ up to second order, that agrees with
the one proposed in \citet{CG} for the case of pure L\'evy
processes. Although we do not study the case of one-dimensional models
in this work, our analysis can be easily adapted to this situation as
well. In particular, the derivation of the pointwise convergence rates
can be directly carried over to this situation.

\subsubsection*{The case of the unknown $ (\sigma_{k})$}
One way to proceed in the case of the unknown $ (\sigma_{k}) $ and $
\nu
_{k}\in\mathfrak{B}_{\gamma} $ with $ \gamma<2 $ is to define $
\widetilde\nu_{k}(x)=x^{4}\nu_{k}(x)$.
Assuming $ \int\widetilde\nu_{k}(x) \,dx<\infty$, we get
\[
\psi^{(4)}_{k}(u)=\int_{\mathbb{R}}e^{\ii ux}\widetilde\nu_{k}(x) \,dx.
\]
Hence, in the above situation
one can apply the regularized Fourier inversion formula to an estimate
of $ \psi^{(4)}_{k}(u) $ instead of $ \psi''_{k}(u)$.

\subsubsection*{Estimation of $ \mathcal{L}_{\Delta}$} Let us first
estimate $ \psi_{k}$. Set
\[
\widehat\psi_{k}(u)=\Delta^{-1}\widehat\phi_{l}(\mathbf{0})\int
_{0}^{u}\frac{\widehat\phi_{k}(v^{(k)})}
{\widehat\phi_{l}(v^{(k)})} \,dv.\vadjust{\goodbreak}
\]
Under Assumptions (AL2), (AT1), (AT2), (AK) and (AH) we derive
%
%
\begin{equation}
\label{psibound}
\|\psi_{k}-\widehat\psi_{k}\|_{L_{\infty}(\mathbb
{R},w)}=O_{\mathrm{a.s}.}\Biggl( \sqrt{\frac{\log^{3+\varepsilon}
n}{n}}\Biggr)
\end{equation}
with a weighting function
\[
w(u)=\biggl[ \int_{0}^{u}\frac{1+|\psi'_{k}(v)|}{|\mathcal
{L}'_{\Delta
}(-\psi_{k}(v))|} \,dv \biggr]^{-1}.
\]
Now let us define an estimate for $ \mathcal{L}_{\Delta} $ as a
solution of the following optimization problem
%
%
\begin{equation}
\label{Lest}
\widehat{\mathcal{L}}_{\Delta}=\arg\inf_{\mathcal{L}\in\mathfrak
{M}_{\Delta}}\sup_{u\in\mathbb{R}}\bigl\{ w(u)\bigl| \mathcal
{L}(-\widehat\psi_{k}(u))-\widehat\phi\bigl(u^{(k)}\bigr) \bigr| \bigr\},
\end{equation}
where $ \mathfrak{M}_{\Delta} $ is the set of completely monotone
functions $ \mathcal{L} $ satisfying $ \mathcal{L}(0)=1 $ and $
\mathcal
{L}'(0)=-\Delta$. Simple calculations and the bound (\ref
{psibound}) yield
%
%
\begin{equation}
\label{Lbound}
\sup_{u\in\mathbb{R}}\{ w(u)| \widehat{\mathcal
{L}}_{\Delta
}(-\psi_{k}(u))-\mathcal{L}_{\Delta}(-\psi_{k}(u)) |
\}
=O_{\mathrm{a.s}.}\Biggl( \sqrt{\frac{\log^{3+\varepsilon} n}{n}}\Biggr).
\end{equation}
Since any function $ \mathcal{L} $ from $ \mathfrak{M}_{\Delta} $
has a representation
\[
\mathcal{L}(u)=\int_{0}^{\infty}e^{-u x}\,dF(x)
\]
with some distribution function $ F $ satisfying $ \int x
\,dF(x)=\Delta
$, we can replace the optimization
over $ \mathfrak{M} $ in (\ref{Lest}) by the optimization over the
corresponding set of distribution functions. The advantage of the
latter approach is that herewith we can directly get an estimate for
the distribution function of the r.v. $ \mathcal{T}(\Delta)$. A
practical implementation of the estimate (\ref{Lest}) is still to be
worked out, as the optimization over the set $ \mathfrak{M}_{\Delta} $
is not feasible and should be replaced by the optimization over
suitable approximation classes (sieves). Moreover, the ``optimal''
weights in (\ref{Lest}) depend on the unknown~$ \mathcal{L}$. However, it turns out that it is possible to use any
weighting function which is dominated by $ w(u)$, that is, one needs only
some lower bounds for~$\mathcal{L}'_{\Delta}$.
\begin{rem}
\label{compfuncbounds}
It is interesting to compare (\ref{psibound}) and (\ref{Lbound}) with
Theorem~3.2 in
\citet{HM}. At first sight it may seem strange that, while the
rates of convergence for our ``link'' function $ \mathcal{L}_{\Delta} $
and the ``components'' $ \psi_{k} $ depend on
the tail behavior of $ \mathcal{L}_{\Delta}'$, the rates in
\citet{HM} rely only on the smoothness of
the link function and the components. The main reason for this is that
the derivative of the link function in the above paper is assumed to be
uniformly bounded from below [assumption (A8)], a restriction that can
be hardly justifiable in our setting. The convergence analysis in the
unbounded case is, in our opinion, an important contribution of this
paper to the problem of estimating composite functions that can be
carried over to other setups and settings.
\end{rem}

%
\subsection{Discussion}
\label{DISC}
As can be seen, the estimate $ \widehat\nu_{k} $ can exhibit various
asymptotic behavior
depending on the underlying L\'evy process $ L_{t} $ and the
time-change $ \mathcal{T}(t)$.
In particular, if the Laplace transform $ \mathcal{L}_{t}(z) $ of $
\mathcal{T} $ dies off at exponential rate as $ \Ree z\to+\infty$ and
$ \mu_{k}=0 $, then the rates of convergence of $ \widehat\nu_{k} $
are logarithmic and depend on the Blumenthal--Geetor index of
the L\'evy process $ L_{t}$. The larger is the Blumenthal--Geetor
index, the slower are the rates and
the more difficult the estimation problem becomes. For the polynomially
decaying $ \mathcal{L}_{t}(z) $ one gets polynomial convergence rates
that also depend on the Blumenthal--Geetor index of $ L_{t} $.
Let us also note that the uniform rates of convergence are usually
rather slow, since $ \beta<1-\gamma$
in most situations. The pointwise convergence rates for points
$ x_{0}\neq0 $ can, on the contrary, be very fast.
The rates obtained turn out to be optimal up to a logarithmic factor in
the minimax sense over the classes $ \mathfrak{S}_{\beta}\cap
\mathfrak
{B}_{\gamma} $ and $ \mathfrak{H}_{s}(x_{0},\delta,D)\cap\mathfrak
{B}_{\gamma}$.

\section{Simulation study}
\label{SIM}
In our simulation study, we consider two models based on time-changed
normal inverse Gaussian (NIG) L\'evy processes.
The NIG L\'evy processes is a relatively new
class of processes introduced in \citet{BN} as a model for log
returns of stock prices.
The processes of this type are characterized by the property that their
increments have NIG distribution. \citet{BN} considered classes of
normal variance--mean
mixtures and defined the NIG distribution as the case when the
mixing distribution is inverse Gaussian.
Shortly after its introduction, it was shown that the NIG
distribution fits very well the log returns on German stock market
data, making the NIG L\'evy processes of great interest for
practioneers. A NIG distribution has in general four parameters: $
\alpha\in\mathbb{R}_{+}$, $ \varkappa\in\mathbb{R}$,
$\delta\in\mathbb{R}_{+} $ and $ \mu\in\mathbb{R} $ with $
|\varkappa
|<\alpha$. Each parameter in $ \operatorname{NIG}(\alpha, \varkappa,
\delta,\mu) $ distribution can be interpreted
as having a different effect on the shape of the distribution: $ \alpha
$ is responsible for the tail heaviness of steepness, $ \varkappa$ has
to do with symmetry, $ \delta$ scales the distribution and $ \mu$
determines its mean value. The NIG distribution is infinitely divisible
with c.f.
\[
\phi(u)=\exp\bigl\{ \delta\bigl( \sqrt{\alpha^{2}-\varkappa
^{2}}-\sqrt
{\alpha^{2}-(\varkappa+\ii u)^{2}}+\ii\mu u \bigr) \bigr\}.
\]
Therefore, one can define the NIG L\'evy process $ (L_{t})_{t\geq0} $ which
starts at zero and has independent and stationary increments such that
each increment $ L_{t+\Delta}-L_{t} $ has $ \operatorname{NIG}(\alpha,
\varkappa, \Delta\delta,\Delta\mu) $ distribution.
The NIG process has no diffusion component making it a pure jump
process with the L\'evy density
%
%
\begin{equation}
\label{NIGNU}
\nu(x)=\frac{2\alpha\delta}{\pi}\frac{\exp(\varkappa
x)K_{1}(\alpha|x|)}{|x|},
\end{equation}
where $ K_{\lambda}(z) $ is the modified Bessel function of the third
kind. Taking into account the asymptotic relations
\[
K_{1}(z)\asymp2/z,\qquad z\to+0,\quad \mbox{and}\quad K_{1}(z)\asymp\sqrt
{\frac{\pi}{2z}} e^{-z},\qquad z\to+\infty,
\]
we conclude that $ \nu\in\mathfrak{B}_{1} $ and $ \nu\in\mathfrak
{H}_{s}(x_{0},\delta,D) $ for arbitrary large $ s>0 $ and some $\delta
>0, D>0$, if $ x_{0}\neq0$. Moreover,  assumption (AL2) is
fulfilled for any $ p>0$.
Furthermore, the identity
\[
\frac{d^{2}}{du^{2}} \log\phi(u)=-\alpha^2/\bigl(\alpha
^{2}-(\varkappa
+\ii u)^{2}\bigr)^{3/2}
\]
implies $ \nu\in\mathfrak{S}_{2-\delta} $ for arbitrary small $
\delta
>0$. In the next sections are going to study two time-changed NIG
processes: one uses the Gamma process as a~time change and another
employs the integrated CIR processes to model~$ \mathcal{T} $.

\subsection{Time change via a Gamma process}
\label{TCGamma}
Gamma process is a L\'evy process such that its increments have Gamma
distribution, so that
$ \mathcal{T} $ is a pure-jump increasing L\'evy process with the L\'
evy density
\[
\nu_{\mathcal{T}}(x)=\theta x^{-1}\exp(-\lambda x),\qquad x\geq0,
\]
where the parameter $ \theta$ controls the rate of jump arrivals and
the scaling parameter $ \lambda$ inversely controls the jump size.
The Laplace transform of $ \mathcal{T} $ is of the form
\[
\mathcal{L}_{t}(z)=(1+z/\lambda)^{-\theta t},\qquad \Ree z\geq0.
\]
It follows from the properties of the Gamma and the corresponding
inverse Gamma distributions that  assumptions (AT1) and (AT2) are
fulfilled for the Gamma process $ \mathcal{T}$,
provided $ \theta\Delta>2/\gamma$.
Consider now the time-changed L\'evy process $ Y_{t}=L_{\mathcal{T}(t)}
$ where $ L_{t}=(L^{1}_{t},L^{2}_{t},L^{3}_{t}) $ is a
three-dimensional L\'evy process with independent NIG components and $
\mathcal{T} $ is a Gamma process. Note that the process $ Y_{t} $ is a
multidimensional L\'evy process since $ \mathcal{T} $ was itself the
L\'
evy process. Let us be more specific and take the $ \Delta
$-increments of the L\'evy processes $ L^{1}_{t}$,
$ L^{2}_{t} $ and $ L^{3}_{t} $ to have $ \operatorname{NIG}(1, -0.05,
1,-0.5)$, $ \operatorname{NIG}(3, -0.05, 1,-1) $ and $ \operatorname
{NIG}(1, -0.03, 1, 2) $ distributions, respectively. Take also $ \theta
=1 $ and $ \lambda=1 $ for the parameters of the Gamma process $
\mathcal{T}$. Next, fix an equidistant grid on $ [0,10] $ of the
length $ n=1\mbox{,}000 $ and simulate a discretized trajectory of the process
$ Y_{t}$.
Let us stress that the dependence structure between the components of $
Y_{t} $ is rather flexible (although they are uncorrelated) and can be
efficiently controlled by the parameters of the corresponding Gamma
process $ \mathcal{T}$.
Next, we construct an estimate $ \widehat\nu_{1} $ as described in
Section \ref{ALG}.
We first estimate the derivatives $ \phi_{1}$, $ \phi_{2}$, $ \phi_{11}
$ and $ \phi_{12} $ by means of (\ref{PhiDeriv1Est}) and~(\ref
{PhiDeriv2Est}). Then we estimate $ \psi''_{1}(u) $ using the formula
(\ref{PsiDeriv2Est1}) with $ k=1 $ and $ l=2$. Finally, we get $
\widehat\nu_{1} $ from (\ref{NUEST}) where the kernel $ \mathcal{K} $
is chosen to be the so-called flat-top kernel of the form
\[
\mathcal{K}(x)=
\cases{
1, &\quad$|x|\leq0.05$, \vspace*{2pt}\cr
\displaystyle \exp\biggl( -\frac{e^{-1/(|x|-0.05)}}{1-|x|} \biggr), &\quad
$0.05<|x|<1$,\cr
0, &\quad$|x|\geq1$.}
\]
The flat-top kernels obviously satisfy  assumption (AK).
Thus, all assumptions of Theorem \ref{UpperBounds} are fulfilled and
Corollary \ref{UPPERCOR2} leads to the following convergence rates for
the estimate $ \widehat\nu_{1} $ of the function $ \bar\nu
_{1}(x)=x^{2}\nu(x) $:
\[
\|\bar\nu_{1}-\widehat\nu_{1}\|_{L_\infty(\mathbb{R},w)}
=O_{\mathrm{a.s}.}
\bigl(n^{-({1-\delta'})/{(\theta\Delta+5/2)}}\log^{({3+\epsilon
'})/{(\theta\Delta+5/2)}}(n) \bigr),\qquad n\to\infty,
\]
with arbitrary small positive numbers $ \delta' $ and $ \epsilon'$,
provided the sequence
$ h_{n} $ is chosen as in Corollary \ref{UPPERCOR2}.
Let us turn to the finite sample performance of the estimate $ \widehat
\nu_{1}$.
It turns out that the choice of the sequence
$ h_{n} $ is crucial for a good performance of~$ \nu_{1} $. For this choice,
we adopt the so called ``quasi-optimality'' approach proposed in
\citet{BR}. This approach is aimed to perform a model selection in
inverse problems without taking into account the noise level.
Although one can prove the optimality of this criterion on average
only, it leads in many situations to quite reasonable results. In order
to implement the ``quasi-optimality'' algorithm in our situation, we
first fix a sequence of bandwidths $ h_{1},\ldots, h_{L} $ and
construct the estimates $ \nu^{(1)}_{1},\ldots, \nu_{1}^{(L)} $ using
the formula (\ref{NUEST}) with bandwidths
$ h_{1},\ldots, h_{L}$, respectively. Then one finds
$ l^{\star}=\argmin_{l} f(l) $ with
\[
f(l)=\bigl\| \widehat\nu_{1}^{(l+1)}- \widehat\nu_{1}^{(l)} \bigr\|
_{L_{1}(\mathbb{R})},\qquad l=1,\ldots,L.
\]
Denote by $ \widetilde\nu_{1}=\widehat\nu^{l^{*}}_{1} $ a new
adaptive estimate for $ \bar\nu_{1}$. In our implementation of the
``quasi-optimality'' approach, we take $ h_{l}=0.5+0.1\times l$, $
l=1,\ldots, 40$.
%
%
\begin{figure}[b]

\includegraphics{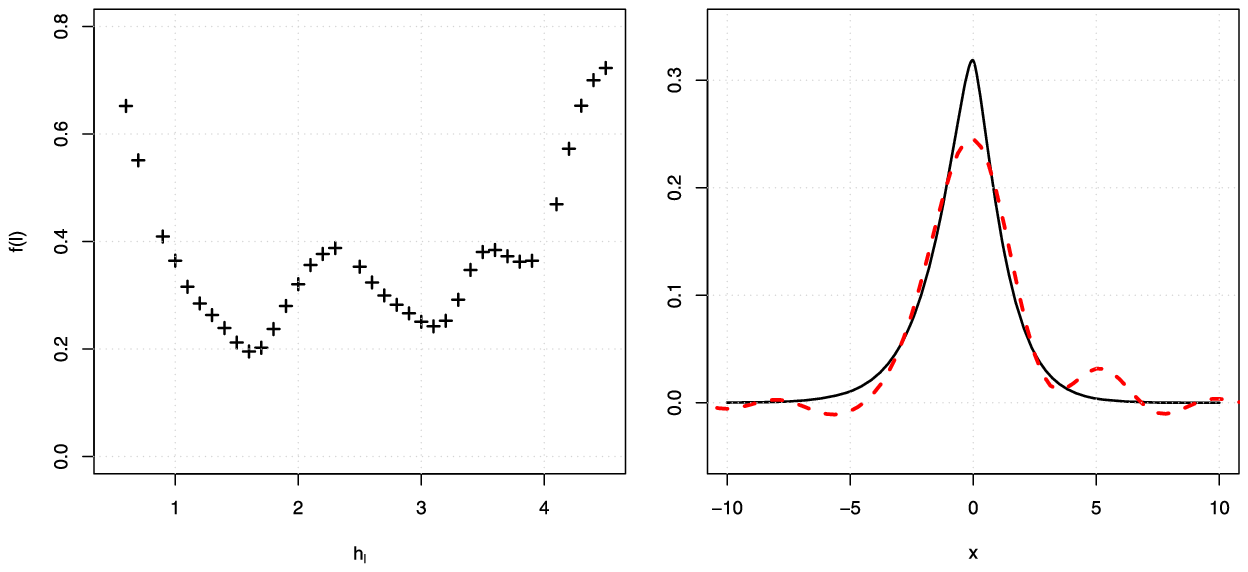}

\caption{Left-hand side: objective function $ f(l)$ for
``quasi-optimality'' approach versus the corresponding bandwidths $
h_{l}$, $l=1,\ldots, 40 $. Right-hand side: adaptive estimate $
\widetilde\nu_{1} $ (dashed line) together with the true function $
\bar\nu_{1}$ (solid line).}
\label{NIGGammaNUEst}%
\end{figure}
In Figure~\ref{NIGGammaNUEst}, the sequence $ f(l)$, $l=1,\ldots,
40$, is plotted. On the right-hand side of
Figure \ref{NIGGammaNUEst}, we show the resulting estimate $
\widetilde\nu_{1} $ together with the true function $ \bar\nu_{1}$.
Based on the estimate $ \widetilde\nu_{1}$, one can estimate some
functionals of~$\bar\nu_{1}$. For example, we have $ \int
\widetilde
\nu_{1}(x) \,dx=1.049053 $ [$ \int\bar\nu_{1}(x) \,dx=1.015189 $].

\subsection{Time change via an integrated CIR process}\label{sec52}
\begin{figure}

\includegraphics{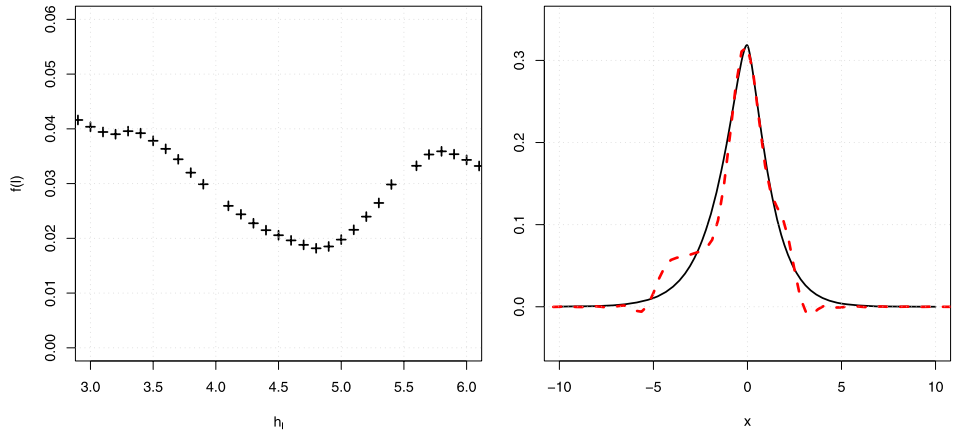}

\caption{Left-hand side: objective function $ f(l)$ for the
``quasi-optimality'' approach versus the corresponding bandwidths $
h_{l} $. Right-hand side: adaptive estimate $ \widetilde\nu_{1} $
(dashed line) together with the true function $ \bar\nu_{1}$ (solid line).}
\label{NIGCIRNUEst}%
\end{figure}

Another possibility to construct a time-changed L\'evy process from the
NIG L\'evy process $ L_{t} $
is to use a time change of the form (\ref{TCProcess}) with some rate
process $ \rho(t)$. A possible candidate for the rate of the time
change is given by the Cox--Ingersoll--Ross process (CIR process). The
CIR process is defined as a solution of the following SDE:
\[
dZ_{t} = \kappa(\eta-Z_{t}) \,dt + \zeta\sqrt{Z_{t}} \,dW_{t},\qquad
Z_{0}=1,
\]
where $ W_{t} $ is a Wiener process.
This process is mean reverting with $ \kappa>0 $ being the speed of
mean reversion, $ \eta>0 $ being the long-run mean rate and $ \zeta>0 $
controlling the volatility of $ Z_{t} $. Additionally, if $ 2\kappa
\eta
>\zeta^{2} $ and $ Z_{0} $ has Gamma distribution, then $ Z_{t} $ is
stationary and exponentially $ \alpha$-mixing [see, e.g., \citet{MH}].
The time change $ \mathcal{T} $ is then defined as
\[
\mathcal{T}(t)=\int_{0}^{t}Z_{t} \,dt.
\]
Simple calculations show that the Laplace transform of $ \mathcal{T}(t)
$ is given by
\[
\mathcal{L}_{t}(z)=\frac{\exp(\kappa^{2}\eta t/\zeta^{2})\exp
(-2z/(\kappa+\gamma(z)\coth(\gamma(z)t/2)))}{(\cosh(\gamma
(z)t/2)+\kappa\sinh(\gamma(z)t/2)/\gamma(z) )^{2\kappa\eta/ \zeta
^{2} }}
\]
with $ \gamma(z)=\sqrt{\kappa^{2}+2\zeta^{2}z}$. It is easy to see
that $ \mathcal{L}_{t}(z)\asymp\exp( -\frac{\sqrt{2z}}{\zeta
}[1+t\kappa\eta] ) $ as $ |z|\to\infty$ with $ \Ree z \geq0$.
Moreover, it can be shown that $ \E|\mathcal{T}(t)|^{p}<\infty$ for
any $ p\in\mathbb{R}$.
Let $ L_{t} $ be again a three-dimensional NIG L\'evy process with
independent components distributed as in Section \ref{TCGamma}.
Construct the time-changed process $ Y_{t}=L_{\mathcal{T}(t)}$. Note
that the process $ Y_{t} $ is not any longer a~L\'evy process and has
in general dependent increments. Let us estimate $ \bar\nu_{1} $, the
transformed L\'evy density of the first component of $ L_{t}$. First,
note that according to Theorem \ref{UpperBounds}, the estimate $
\widehat\nu_{1} $ constructed as described in Section~\ref{ALG}, has
the following logarithmic convergence rates
\[
\|\bar\nu_{1}-\widehat\nu_{1}\|_{L_\infty(\mathbb{R},w)}
=O_{\mathrm{a.s}.}
\bigl(\log^{-2(2-\delta)} (n) \bigr),\qquad n\to\infty,
\]
for arbitrary small $ \delta>0$, provided the bandwidth sequence is
chosen in the optimal way. Finite sample performance of $ \widehat\nu
_{1} $ with the choice of $ h_{n} $ based on the ``quasi-optimality''
approach is illustrated in
Figure \ref{NIGCIRNUEst} where the sequence of estimates $ \widehat
\nu^{(1)}_{1},\ldots, \widehat\nu^{(L)}_{1} $ was constructed from the
time series $ Y_{\Delta},\ldots, Y_{n\Delta} $ with $ n=5\mbox{,}000 $
and $ \Delta=0.1$.
%
%
The parameters of the used CIR process are $ \kappa=1 $, $ \eta=1 $ and
$ \zeta=0.1$. Again we can compute some functionals of $ \widetilde
\nu
_{1}$. We have, for example, following estimates for the integral and
for the mean of $ \bar\nu_{1}$:
$ \int\widetilde\nu_{1}(x) \,dx=1.081376 $ [$ \int\bar\nu_{1}(x)
\,dx=1.015189 $] and $ \int x\widetilde\nu_{1}(x) \,dx=-0.4772505 $ [$
\int x\bar\nu_{1}(x) \,dx=-0.3057733 $].

\begin{figure}

\includegraphics{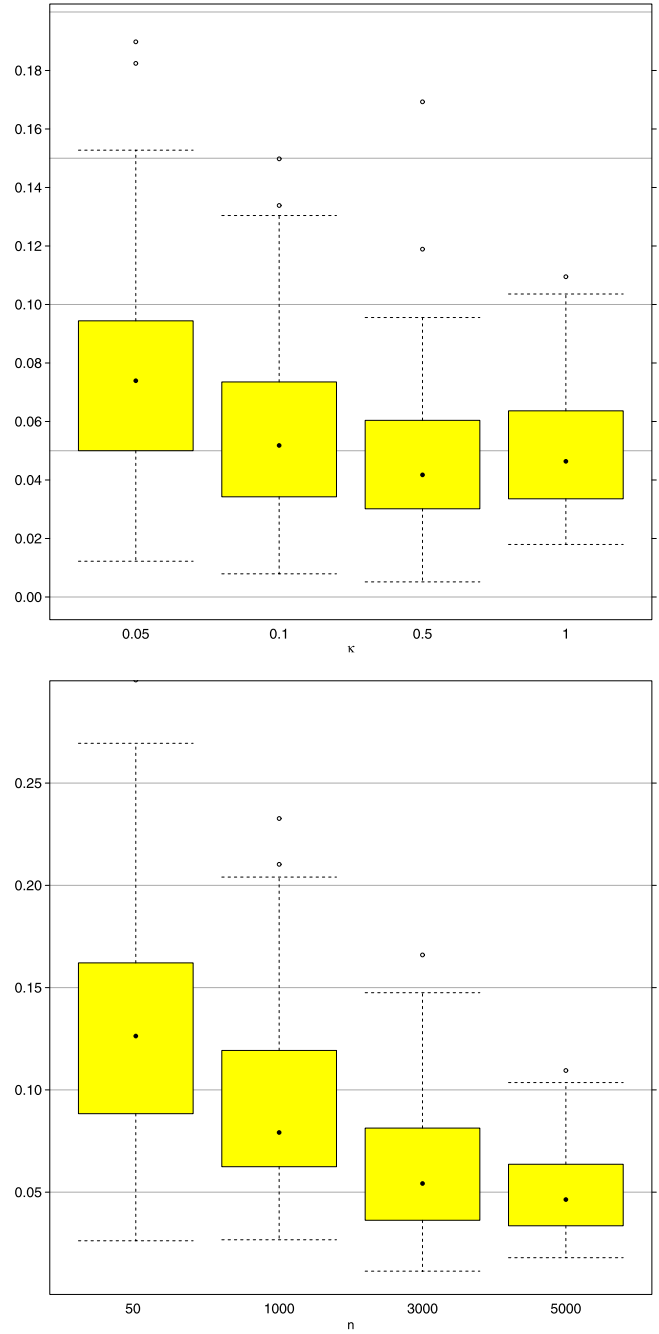}

\caption{Boxplots of the error
$\|\bar\nu_{1}-\widetilde\nu_{1}\|_{L_\infty(\mathbb{R},w)} $ for
different values of the mean reversion speed parameter $ \kappa$ and
different numbers of observations $ n$. } \label{boxplots}
\end{figure}

Let us now test the performance of estimation algorithm in the case of
a time-changed NIG process (parameters are the same as before), where
the time change is again given by the integrated CIR process with the
parameters $ \eta=1$, $\zeta=0.1 $ and $ \kappa\in\{0.05,0.1,0.5,1\}$.
Figure \ref{boxplots}(left) shows the boxplots of the resulting error $
\|\bar\nu_{1}-\widetilde\nu_{1}\|_{L_\infty(\mathbb{R},w)} $
computed using $ 100 $ trajectories each of the length $ n=5\mbox{,}000$,
where the time span between observation is $ \Delta=0.1$.
%
%
Note that if our time units are days, then we get about two years of
observations with about one mean reversion per month in the case $
\kappa=0.05$. As one can see, the performance of the algorithm remains
reasonable for the whole range of $ \kappa$.
In Figure \ref{boxplots}(right), we present the boxplots of the error $
\|\bar\nu_{1}-\widetilde\nu_{1}\|_{L_\infty(\mathbb{R},w)} $
in the case of $ \eta=1$, $\zeta=0.1$, $ \kappa=1$ and $ n\in\{500,
1\mbox{,}000, 3\mbox{,}000, 5\mbox{,}000 \}$. As one can expect, the performance of the
algorithm becomes worse as $ n $ decreases. However,
the quality of the estimation remains reasonable even for $ n=500$.

\section{Proofs of the main results}

\subsection{\texorpdfstring{Proof of Theorem \protect\ref{UpperBounds}}{Proof of Theorem 4.4}}
For simplicity, let consider the case of $ \mu_{l}>0 $ and $ \sigma
_{k}=0$. By Proposition \ref{ExpBounds}
[take $ G_{n}(u,z)=\exp(\ii u z)$, $ L_{n}=\bar\mu_{n}=\bar\sigma_{n}=1
$, $ a=0, b=1 $]
\[
\mathbb{P}\bigl(|\widehat\phi_{l}(\mathbf{0})|\leq\kappa/\sqrt{n}
\bigr)\geq
\mathbb{P}\bigl(|\widehat\phi_{l}(\mathbf{0})-\phi_{l}(\mathbf{0})|> \mu
_{l}\bigr)\leq B n^{-1-\delta}
\]
for some constants $ \delta>0, B>0 $ and $ n $ large enough.
Furthermore, simple calculations lead to the following representation:
%
%
\begin{eqnarray}
\label{psirepr}
\psi_{k}^{\prime\prime}(u)-\widehat{\psi}_{k,2}(u) &=&\frac{\psi
''_{k}(u)}{\psi'_{l}(0)}\bigl(
\phi_{l}(\mathbf{0})-\widehat{\phi}_{l}(\mathbf{0})
\bigr)\nonumber\\[-8pt]\\[-8pt]
&&{}+\mathcal{R}_{0}(u)+
\mathcal{R}_{1}(u)+\mathcal{R}_{2}(u),\nonumber
\end{eqnarray}
where
\begin{eqnarray*}
\mathcal{R}_{0}(u)&=&[V_{1}(u)\psi''_{k}(u)-V_{2}(u)\psi'_{k}(u)
]\bigl(
\phi_{l}\bigl(u^{(k)}\bigr)-\widehat{\phi}_{l}\bigl(u^{(k)}\bigr)\bigr) \\
&&{}+V_{2}(u)\bigl( \phi_{k}\bigl(u^{(k)}\bigr)-\widehat{\phi
}_{k}\bigl(u^{(k)}\bigr)\bigr)
\\
&&{}-V_{1}(u)\bigl( \phi_{kk}\bigl(u^{(k)}\bigr)-\widehat{\phi
}_{kk}\bigl(u^{(k)}\bigr)\bigr) \\
&&{}+V_{1}(u)\psi'_{k}(u)\bigl( \phi_{lk}\bigl(u^{(k)}\bigr)-\widehat{\phi
}_{lk}\bigl(u^{(k)}\bigr)\bigr),
\\
\mathcal{R}_{1}(u)&=&[\widetilde V_{1}(u)\psi
''_{k}(u)-\widetilde
V_{2}(u)\psi'_{k}(u) ]\bigl(
\phi_{l}\bigl(u^{(k)}\bigr)-\widehat{\phi}_{l}\bigl(u^{(k)}\bigr)\bigr) \\
&&{}+\widetilde V_{2}(u)\bigl( \phi_{k}\bigl(u^{(k)}\bigr)-\widehat{\phi
}_{k}\bigl(u^{(k)}\bigr)\bigr)
\\
&&{}-\widetilde V_{1}(u)\bigl( \phi_{kk}\bigl(u^{(k)}\bigr)-\widehat{\phi
}_{kk}\bigl(u^{(k)}\bigr)\bigr) \\
&&{}+\widetilde V_{1}(u)\psi'_{k}(u)\bigl( \phi_{lk}\bigl(u^{(k)}\bigr)-\widehat
{\phi}_{lk}\bigl(u^{(k)}\bigr)\bigr),
\\
\mathcal{R}_{2}(u) &=&\Gamma^{2}(u)\frac{\phi_{l}(\mathbf{0})(
\phi
_{lk}(u^{(k)})-\widehat{\phi}_{lk}(u^{(k)})) }{[ \phi
_{l}(u^{(k)})] ^{2}}\\
&&{}\times\bigl[ \bigl( \phi_{l}\bigl(u^{(k)}\bigr)-\widehat
{\phi}%
_{l}\bigl(u^{(k)}\bigr)\bigr) \psi_{k}^{\prime}(u)-\bigl( \phi_{k}\bigl(u^{(k)}\bigr)-%
\widehat{\phi}_{k}\bigl(u^{(k)}\bigr)\bigr) \bigr]\\
&&{} +\frac{(\widehat{\phi
}_{l}(\mathbf{0})-\phi_{l}(\mathbf{0}))}{%
\phi_{l}(u^{(k)})}\biggl[\frac{\mathcal{R}_{0}+\mathcal
{R}_{1}}{\phi
_{l}(\mathbf{0})}\biggr]
\end{eqnarray*}
with%
\begin{eqnarray*}
V_{1}(u) &=&\frac{\phi_{l}(\mathbf{0})}{\Delta\phi_{l}(u^{(k)})}=
-\frac{1}{\mathcal{L}_{\Delta}^{\prime}(-\psi_{k}(u))},
\\
V_{2}(u) &=&\frac{\phi_{l}(\mathbf{0})\phi_{lk}(u^{(k)})}{\Delta
[ \phi_{l}(u^{(k)})] ^{2}}=-V_{1}(u)\psi'_{k}(u)\frac
{\mathcal{L}%
_{\Delta}^{\prime\prime}(-\psi_{k}(u))}{\mathcal{L}_{\Delta
}^{\prime
}(-\psi
_{k}(u))},
\\
\widetilde V_{1}(u)&=&\bigl(\Gamma(u)-1\bigr)V_{1}(u),\qquad \widetilde
V_{2}(u)=\bigl(\Gamma^{2}(u)-1\bigr)V_{2}(u)
\end{eqnarray*}
and
\[
\Gamma(u)=\biggl[ 1-\frac{1}{\phi_{l}(u^{(k)})}\bigl( \phi
_{l}\bigl(u^{(k)}\bigr)-%
\widehat{\phi}_{l}\bigl(u^{(k)}\bigr)\bigr) \biggr] ^{-1}.
\]
The representation (\ref{psirepr}) and the Fourier inversion formula
imply the following representation for
the deviation $ \bar\nu_{k}-\widehat\nu_{k} $:
\begin{eqnarray*}
\bar\nu_{k}(x)-\widehat\nu_{k}(x)&=&\frac{1}{2\pi}\frac{(
\phi_{l}(\mathbf{0})-\widehat{\phi}_{l}(\mathbf{0}))}{\psi
'_{l}(0)}\int_{\mathbb{R}} e^{-\ii u x}\psi''_{k}(u)\mathcal
{K}(uh_{n}) \,du\\
&&{}+\frac{1}{2\pi}\int_{\mathbb{R}} e^{-\ii u
x}\mathcal
{R}_{0}(u)\mathcal{K}(uh_{n}) \,du
\\
&&{} +\frac{1}{2\pi}\int_{\mathbb{R}} e^{-\ii u x}\mathcal
{R}_{1}(u)\mathcal{K}(uh_{n}) \,du\\
&&{}+\frac{1}{2\pi}\int_{\mathbb{R}} e^{-\ii u x}\mathcal
{R}_{2}(u)\mathcal
{K}(uh_{n}) \,du
\\
&&{}+\frac{1}{2\pi}\int_{\mathbb{R}} e^{-\ii u x}\bigl(1-\mathcal
{K}(uh_{n})\bigr)\psi''_{k}(u) \,du.
\end{eqnarray*}
First, let us show that
\[
\sup_{x\in\mathbb{R}}\biggl| \int_{\mathbb{R}}e^{-\ii u x}\mathcal
{R}_{1}(u)\mathcal{K}(uh_{n}) \,du \biggr|=o_{\mathrm{a.s}}\Biggl( \sqrt
{\frac
{\log^{3+\varepsilon} n}{n}\int_{-1/h_{n}}^{1/h_{n}}\mathfrak
{R}^{2}_{k}(u) \,du} \Biggr)
\]
and
\[
\sup_{x\in\mathbb{R}}\biggl| \int_{\mathbb{R}}e^{-\ii u x}\mathcal
{R}_{2}(u)\mathcal{K}(uh_{n}) \,du \biggr|=o_{\mathrm{a.s}}\Biggl( \sqrt
{\frac
{\log^{3+\varepsilon} n}{n}\int_{\mathbb{R}}\mathfrak
{R}^{2}_{k}(u)
\,du} \Biggr).
\]
We have, for example, for the first term in $ \mathcal{R}_{1}(u) $
\begin{eqnarray*}
&&\biggl| \int_{\mathbb{R}}e^{-\ii u z} \bigl(\Gamma(u)-1\bigr)V_{1}(u)\psi
''_{k}(u)\bigl(
\phi_{l}\bigl(u^{(k)}\bigr)-\widehat{\phi}_{l}\bigl(u^{(k)}\bigr)\bigr)\mathcal
{K}(uh_{n}) \,du \biggr|
\\
&&\qquad\leq\sup_{|u|\leq1/h_{n}}|\Gamma(u)-1|\sup_{u\in\mathbb{R}}\bigl[
w(|u|)\bigl|\phi_{l}\bigl(u^{(k)}\bigr)-\widehat{\phi}_{l}\bigl(u^{(k)}\bigr)\bigr|
\bigr]w^{-1}(1/h_{n})\\
&&\qquad\quad{}\times\int_{-1/h_{n}}^{1/h_{n}}|V_{1}(u)||\psi''_{k}(u)| \,du
\end{eqnarray*}
with $ w(u)=\log^{-1/2}(e+u), u\geq0$.
Fix some $ \xi>0 $ and consider the event
\[
\mathcal{A}=\Biggl\{ \sup_{\{|u|\leq1/h_{n}\}}\bigl[
w(|u|)\bigl|\widehat
\phi_{l}\bigl(u^{(k)}\bigr)-\phi_{l}\bigl(u^{(k)}\bigr)\bigr| \bigr]
\leq\xi\sqrt{\frac
{\log
n}{n}}\Biggr\}.
\]
By  assumption (AH), it holds on $ \mathcal{A} $ that
\begin{eqnarray*}
\sup_{|u|<1/h_{n}}\biggl| \frac{\phi_{l}(u^{(k)})-%
\widehat{\phi}_{l}(u^{(k)})}{\phi_{l}(u^{(k)})} \biggr|&\leq&\xi
M_{n}w^{-1}(1/h_{n})\sqrt{\log n/n}\\
&=&o\bigl(\sqrt{h_{n}}\bigr),\qquad
n\to\infty,
\end{eqnarray*}
and hence
%
%
\begin{equation}
\label{GammaEst}
\sup_{\{|u|\leq1/h_{n}\}}|1-\Gamma(u)|=o\bigl(\sqrt{h_{n}}
\bigr),\qquad n\to\infty.
\end{equation}
Therefore, one has on $ \mathcal{A} $ that
\begin{eqnarray*}
\sup_{x\in\mathbb{R}}\biggl| \int_{-1/h_{n}}^{1/h_{n}}e^{-\ii u x}
\bigl(\Gamma(u)-1\bigr)V_{1}(u)\psi''_{k}(u)\bigl(
\phi_{l}\bigl(u^{(k)}\bigr)-\widehat{\phi}_{l}\bigl(u^{(k)}\bigr)\bigr)\mathcal
{K}(uh_{n}) \,du \biggr|
\\
= o\Biggl( \sqrt{\frac{h_{n}\log^{2} n}{n}}\int
_{-1/h_{n}}^{1/h_{n}}\mathfrak{R}_{k}(u) \,du \Biggr)
= o\Biggl( \sqrt{\frac{\log^{3+\varepsilon} n}{n}\int
_{-1/h_{n}}^{1/h_{n}}\mathfrak{R}^{2}_{k}(u) \,du} \Biggr)
\end{eqnarray*}
since $ \psi''_{k}(u) $ and $ \mathcal{K}(u) $ are uniformly bounded on
$ \mathbb{R}$.
On the other hand, Proposition~\ref{ExpBounds}
implies [on can take $ G_{n}(u,z)=\exp(\ii u z)$, $ L_{n}=\bar\mu
_{n}=\bar\sigma_{n}=1 $, $ a=0$, \mbox{$b=1 $}]
\[
\mathbb{P}(\bar{\mathcal{A}})\lesssim n^{-1-\delta'},\qquad n\to\infty,
\]
for some $ \delta'>0$.
The Borel--Cantelli lemma yields
\begin{eqnarray*}
&&\sup_{x\in\mathbb{R}}\biggl| \int_{-1/h_{n}}^{1/h_{n}}e^{-\ii u x}
\bigl(\Gamma(u)-1\bigr)V_{1}(u)\psi''_{k}(u)\bigl(
\phi_{l}\bigl(u^{(k)}\bigr)-\widehat{\phi}_{l}\bigl(u^{(k)}\bigr)\bigr)\mathcal
{K}(uh_{n}) \,du \biggr|
\\
&&\qquad=o_{\mathrm{a.s}.}\Biggl( \sqrt{\frac{\log^{3+\varepsilon} n}{n}\int
_{-1/h_{n}}^{1/h_{n}}\mathfrak{R}^{2}_{k}(u) \,du} \Biggr).
\end{eqnarray*}
Other terms in $ \mathcal{R}_{1} $ and $ \mathcal{R}_{2} $ can be
analyzed in a similar way.
Turn now to the rate determining term $ \mathcal{R}_{0}$. Consider,
for instance, the integral
%
%
\begin{eqnarray}\qquad
\label{R0}
&&\int_{-1/h_{n}}^{1/h_{n}}e^{-\ii u x}V_{1}(u)\psi''_{k}(u)\bigl(
\phi_{l}\bigl(u^{(k)}\bigr)-\widehat{\phi}_{l}\bigl(u^{(k)}\bigr)\bigr)\mathcal
{K}(uh_{n}) \,du
\nonumber\\[-8pt]\\[-8pt]
&&\qquad=\frac{1}{nh_{n}}\sum_{j=1}^{n}\biggl[ Z_{j}^{l}K_{n}\biggl( \frac
{x-Z_{j}^{k}}{h_{n}}%
\biggr) -\E\biggl\{ Z^{l}\frac{1}{h_{n}}K_{n}\biggl(
\frac{x-Z^{k}}{h_{n}}\biggr) \biggr\} \biggr]
=\mathcal{S}(x)
\nonumber
\end{eqnarray}
with
\[
K_{n}(z)=\int_{-1}^{1}e^{-\ii uz} V_{1}(u/h_{n})\psi
_{k}^{\prime\prime}(u/h_{n})\mathcal{K}(u) \,du.
\]
Now we are going to make use of Proposition \ref{ExpBounds} to estimate
the term $ \mathcal{S}(x) $ on the r.h.s. of (\ref{R0}). To this end,
let
\[
G_{n}(u,z)= \frac{1}{h_{n}}K_{n}\biggl( \frac{u-z}{h_{n}}\biggr).
\]
Since $ \nu_{k},\nu_{l}\in\mathfrak{B}_{\gamma} $ for some $
\gamma>0
$ [assumption (AL1)],
the L\'evy processes $ L^{k}_{t} $ and $ L^{l}_{t} $ possess infinitely
smooth densities $ p_{k,t} $ and $ p_{l,t} $ which are bounded for $
t>0 $
[see \citet{SA}, Section 28] and fulfill [see \citet{P}]
%
%
\begin{eqnarray}
\label{pka}
\sup_{x\in\mathbb{R}} \{ p_{k,t}(x) \}&\lesssim&
t^{-1/\gamma},\qquad t\to0,
\\
\label{pla}
\sup_{x\in\mathbb{R}} \{ p_{l,t}(x) \}&\lesssim&
t^{-1/\gamma},\qquad t\to0.
\end{eqnarray}
Moreover, under assumption (AL2) [see \citet{LP}]
%
%
\begin{equation}\quad
\label{mpa1}
\int|x|^{m} p_{k,t}(x) \,dx = O(t),\qquad \int|x|^{m} p_{l,t}(x)
\,dx = O(t),\qquad t\to0,
\end{equation}
and
%
%
\begin{eqnarray}
\label{mpa2}
\int|x|^{m} p_{k,t}(x) \,dx &=& O(t^{m}),\nonumber\\[-8pt]\\[-8pt]
\int|x|^{m}p_{l,t}(x) \,dx &=& O(t^{m}),\qquad t\to+\infty,\nonumber
\end{eqnarray}
for any $ 2\leq m\leq p$.
As a result, the distribution of $ (Z^{k},Z^{l}) $ is absolutely
continuous with uniformly bounded density $ q_{kl} $ given by
\[
q_{kl}(y,z)=\int_{0}^{\infty} p_{k,t}(y)p_{l,t}(z) \,d\pi(dt),
\]
where $ \pi$ is the distribution function of the r.v. $\mathcal
{T}(\Delta)$.
The asymptotic relations (\ref{pka})--(\ref{mpa2}) and  assumption
(AT1) imply
\begin{eqnarray*}
\E[| Z^{l} |^{2}|G_{n}(u,Z^{k})|^{2}]&=&
\frac
{1}{h^{2}_{n}}\int_{\mathbb{R}}\biggl| K_{n}\biggl( \frac
{u-y}{h_{n}}\biggr)\biggr|^{2}\biggl\{ \int_{\mathbb
{R}}|z|^{2}q_{kl}(y,z) \,dz \biggr\} \,dy
\\
&\leq& \frac{C_{0}}{h_{n}}\int_{\mathbb{R}}| K_{n}(v)
|^{2} \,dv
\\
&\leq& C_{1}\int_{-1/h_{n}}^{1/h_{n}}|V_{1}(u)|^{2} \,du
\end{eqnarray*}
with some finite constants $ C_{0}>0 $ and $ C_{1}>0$.
Similarly,
\begin{eqnarray*}
\E[|Z^{k}|^{2}|G_{n}(u,Z^{k})|^{2}]& \leq&
C_{2}\int_{-1/h_{n}}^{1/h_{n}}|V_{1}(u)|^{2} \,du,
\\
\E[|Z^{k}|^{4}|G_{n}(u,Z^{k})|^{2}]& \leq&
C_{3}\int_{-1/h_{n}}^{1/h_{n}}|V_{1}(u)|^{2} \,du,
\\
\E[|Z^{k}|^{2}|Z^{l}
|^{2}|G_{n}(u,Z^{k})|^{2}]& \leq& C_{4}\int
_{-1/h_{n}}^{1/h_{n}}|V_{1}(u)|^{2} \,du
\end{eqnarray*}
with some positive constants $ C_{2}, C_{3} $ and $ C_{4}$.
Define
\begin{eqnarray*}
\bar\sigma^{2}_{n}&=&C\int_{-1/h_{n}}^{1/h_{n}} |V_{1}(u)|^{2} \,du,
\\
\bar\mu_{n}&=&\| \mathcal{K} \|_{\infty}\| \psi'' \|_{\infty}\int
_{-1/h_{n}}^{1/h_{n}} |V_{1}(u)| \,du,
\\
L_{n}&=&\| \mathcal{K} \|_{\infty}\| \psi'' \|_{\infty}\int
_{-1/h_{n}}^{1/h_{n}} |u||V_{1}(u)| \,du,
\end{eqnarray*}
where $ C=\max_{k=1,2,3,4}\{ C_{k} \}$.
Since $ |V_{1}(u)|\to\infty$ as $ |u|\to\infty$ and $ h_{n}\to
\infty, $ we get $ \bar\mu_{n}/\bar\sigma^{2}_{n}=O(1)$. Furthermore,
due to assumption (AH)
%
%
\begin{equation}
\label{ARMS}
\bar\mu_{n}\lesssim h_{n}^{-1/2}\bar\sigma_{n}\lesssim
n^{1/2-\delta
/2} \bar\sigma_{n},\quad L_{n}\lesssim h_{n}^{3/2} \bar\sigma_{n}
\lesssim n^{3/2} \bar\sigma_{n},\quad n\to\infty,\hspace*{-28pt}
\end{equation}
and $ \bar\sigma_{n}=O(h_{n}^{-1/2}M_{n})=O(n^{1/2})$.
Thus,  assumptions (AG1) and (AG2) of Proposition \ref{ExpBounds}
are fulfilled.
  Assumption (AZ1) follows from Lemma \ref{MIX}
and  assumption (AT1).
Therefore, we get by Proposition \ref{ExpBounds}
\[
\mathbb{P}\Biggl( \sup_{z\in\mathbb{R}}[ w(|z|)| \mathcal{S}(z)
| ]\geq\xi\sqrt{\frac{\bar\sigma^{2}_{n}\log
^{3+\varepsilon}n}{n}} \Biggr)\lesssim n^{-1-\delta' }
\]
for some $ \delta'>0 $ and $ \xi>\xi_{0}$. Noting that
\[
\bar\sigma_{n}^{2}\leq C\int_{-1/h_{n}}^{1/h_{n}}\mathfrak
{R}^{2}_{k}(u) \,du,
\]
we derive
\[
\sup_{z\in\mathbb{R}}[ w(|z|)| \mathcal{S}(z) |
]
=O_{\mathrm{a.s}.}
\Biggl( \sqrt{\frac{\log^{3+\varepsilon} n}{n}\int
_{-1/h_{n}}^{1/h_{n}}\mathfrak{R}^{2}_{k}(u) \,du} \Biggr).
\]
Other terms in $ \mathcal{R}_{0} $ can be studied in a similar manner. Finally,
%
%
\begin{eqnarray}
\label{nudev}
\| \widehat\nu_{k}-\bar\nu_{k} \|_{L_{\infty}(\mathbb{R},w)}
&=& O_{\mathrm{a.s}.}
\Biggl( \sqrt{\frac{\log^{3+\varepsilon} n}{n}\int
_{-1/h_{n}}^{1/h_{n}}\mathfrak{R}^{2}_{k}(u) \,du} \Biggr)
\nonumber\\[-8pt]\\[-8pt]
&&{}+
\frac{1}{2\pi}\int_{\mathbb{R}} |1-\mathcal{K}(uh_{n})||\psi
_{k}^{\prime\prime}(u)| \,du.\nonumber
\end{eqnarray}
The second, bias term on the r.h.s. of (\ref{nudev}) can be easily
bounded if we recall that $ \nu_{k}\in\mathfrak{S}_{\beta} $ and $
\mathcal{K}(u)=1 $ on $ [-a_{K},a_{K}] $
\begin{eqnarray*}
\frac{1}{2\pi}\int_{\mathbb{R}} |1-\mathcal{K}(uh_{n})||\psi
_{k}^{\prime\prime}(u)| \,du
&\lesssim& h_{n}^{\beta}\int_{\{|u|>a_{K}/h_{n}\}} |u|^{\beta
}|\mathbf
{F}[\bar\nu_{k}](u)| \,du
\\
&\lesssim& h_{n}^{\beta}\int_{\mathbb{R}} (1+|u|^{\beta})|\mathbf
{F}[\bar\nu_{k}](u)| \,du,\qquad n\to\infty.
\end{eqnarray*}

%
\subsection{\texorpdfstring{Proof of Theorem \protect\ref{pointwiseupper}}{Proof of Theorem 4.7}}
We have
\begin{eqnarray*}
\widehat\nu_{k}(x_{0})-\bar\nu_{k}(x_{0})&=&\biggl[ \frac{1}{2\pi
}\int
_{\mathbb{R}}e^{-\ii ux_{0}}\psi''_{k}(u)\mathcal{K}(uh_{n})
\,du-\bar
\nu_{k}(x_{0}) \biggr]
\\
&&{}+\frac{1}{2\pi}\int_{\mathbb{R}}e^{-\ii ux_{0}}\bigl(\widehat\psi
_{k,2}-\psi''_{k}(u)\bigr)\mathcal{K}(uh_{n}) \,du\\
&=&J_{1}+J_{2}
\end{eqnarray*}
Introduce
\[
K(z)=\frac{1}{2\pi}\int_{-1}^{1}e^{\ii u z} \mathcal{K}(u) \,du,
\]
then by the Fourier inversion formula
%
%
\begin{equation}
\label{KFInv}
\mathcal{K}(u)=\int_{\mathbb{R}}e^{-\ii uz}K(z) \,dz.
\end{equation}
  Assumption (AK) together with the smoothness of $ \mathcal{K} $
implies that $ K(z) $ has finite absolute moments up to order $ m \geq
s$ and it holds that
%
%
\begin{equation}
\label{AKM}
\int K(z) \,dz=1,\qquad \int z^{k}K(z) \,dz=0, \qquad k=1,\ldots,m.
\end{equation}
Hence
\[
J_{1}= \int_{-\infty}^{\infty}\bar\nu_{k}(x_{0}+h_{n}v)K(v)
\,dv-\bar
\nu_{k}(x_{0})
\]
and
\begin{eqnarray*}
| J_{1} |&\leq&
\biggl| \int_{|v|>\delta/h_{n}}[\bar\nu_{k}(x_{0})-\bar\nu
_{k}(x_{0}+h_{n}v)]K(v) \,dv \biggr|
\\
&&{} + \biggl| \int_{|v|\leq\delta/h_{n}}[\bar\nu_{k}(x_{0})-\bar
\nu
_{k}(x_{0}+h_{n}v)]K(v) \,dv \biggr|
\\
&=& I_{1}+I_{2}.
\end{eqnarray*}
Since $ \| \bar\nu\|_{\infty}\leq C_{\bar\nu} $ for some constant $
C_{\bar\nu}>0$, we get
\[
I_{1}\leq2C_{\bar\nu} \int_{|v|>\delta/h_{n}}|K(v)| \,dv\leq
C_{\bar
\nu} C_{K}(h_{n}/\delta)^{m}
\]
with $ C_{K}=\int_{\mathbb{R}}|K(v)||v|^{m} \,dv$.
Further, by the Taylor expansion formula,
\begin{eqnarray*}
I_{2}&\leq& \Biggl| \sum_{j=0}^{s-1}\frac{h_{n}^{j}\bar\nu
_{k}^{(j)}(x_{0})}{j!}\int_{|v|\leq\delta/h_{n}}K(v)v^{j} \,dv \Biggr|
\\
&&{} +\biggl| \int_{|v|\leq\delta/h_{n}} K(v) \biggl[ \int
_{x_{0}}^{x_{0}+h_{n}v} \frac{\bar\nu_{k}^{(s)}(\zeta)(\zeta
-x_{0})^{s-1}}{(s-1)!} \,d\zeta\biggr] \,dv \biggr|
\\
&=&I_{21}+I_{22}.
\end{eqnarray*}
First, let us bound $ I_{21} $ from above. Note that, due to (\ref{AKM}),
\[
I_{21}=\Biggl| \sum_{j=0}^{s-1}\frac{h_{n}^{j}\bar\nu
_{k}^{(j)}(x_{0})}{j!}\int_{|v|> \delta/h_{n}}K(v)v^{j} \,dv \Biggr|.
\]
Hence,
\begin{eqnarray*}
I_{21}&\leq& \biggl( \frac{h_{n}}{\delta} \biggr)^{m} \sum
_{j=0}^{s-1}\frac{\delta^{j}|\bar\nu_{k}^{(j)}(x_{0})|}{j!}\int_{|v|>
\delta/h_{n}}|K(v)||v|^{m} \,dv
\\
&\leq& \biggl( \frac{h_{n}}{\delta} \biggr)^{m} L C_{K}\exp(\delta).
\end{eqnarray*}
Furthermore, we have for $ I_{22} $
\[
I_{22} \leq\frac{L
h_{n}^{s}}{s!}\int_{|v|\leq\delta/ h_{n}}|K(v)||v|^{s} \,dv.
\]
Combining all previous inequalities and taking into account the fact
that $ m\geq s $, we derive
\[
|J_{1}|\lesssim h_{n}^{s},\qquad n\to\infty.
\]
The stochastic term $ J_{2} $ can handled along the same lines as in
the proof of Theorem~\ref{UpperBounds}.

\subsection{\texorpdfstring{Proof of Theorem \protect\ref{LowBounds}}{Proof of Theorem 4.9}}
Define
\[
K_{0}(x)=\prod_{k=1}^{\infty}\biggl( \frac{\sin(a_{k}x)}{a_{k}x}
\biggr) ^{2}
\]
with $ a_{k}=2^{-k}, k\in\mathbb{N}$. Since $ K_{0}(x) $ is
continuous at $ 0 $ and does not vanish there, the function
\[
K(x)=\frac{1}{2\pi}\frac{\sin(2x)}{\pi x}\frac{K_{0}(x)}{K_{0}(0)}
\]
is well defined on $ \mathbb{R}$.
Next, fix two positive numbers $ \beta$ and $ \gamma$ such that $
\gamma\in(0,1) $ and $ 0<\beta<1-\gamma$. Consider a function
\[
\Phi(u)=\frac{e^{\ii x_{0}u}}{(1+u^{2})^{(1+\beta)/2}\log^{2}(e+u^{2})}
\]
for some $x_{0}>0$ and define%
\[
\mu_{h}(x)=\int_{-\infty}^{\infty}\mu(x+zh)K(z) \,dz
\]
for any $ h>0$, where
\[
\mu(x)=\frac{1}{2\pi}\int_{-\infty}^{\infty}e^{-\ii xu}\Phi(u)\,du.
\]
In the next lemma, some properties of the functions $ \mu$ and $ \mu
_{h} $ are collected.
\begin{lem}
\label{MuProp}
Functions $ \mu$ and $ \mu_{h} $ have the following properties:
\begin{longlist}
\item$ \mu$ and $ \mu_{h} $ are uniformly bounded on $ \mathbb
{R}$,
\item for any natural $ n>0 $
%
%
\begin{equation}
\label{MuDecay}
\max\{ \mu(x), \mu_{h}(x) \}\lesssim|x|^{-n},\qquad |x|\to\infty,
\end{equation}
that is, both functions $ \mu(x) $ and $ \mu_{h}(x)$ decay faster than
any negative power of $ x $,
\item it holds
%
%
\begin{equation}
\label{MuMuh}
x_{0}^{2}\mu(x_{0})-x_{0}^{2}\mu_{h}(x_{0})\geq Dh^{\beta}\log^{-1}(1/h)
\end{equation}
for some constant $ D>0 $ and $ h $ small enough.
\end{longlist}
\end{lem}

Fix some $ \varepsilon>0 $ and consider two functions
\begin{eqnarray*}
\nu_{1}(x) &=&\nu_{\gamma}(x)+\frac{1-\varepsilon}{(1+x^{2})^{2}}%
+\varepsilon\mu(x), \\
\nu_{2}(x) &=&\nu_{\gamma}(x)+\frac{1-\varepsilon}{(1+x^{2})^{2}}%
+\varepsilon\mu_{h}(x),
\end{eqnarray*}
where $\nu_{\gamma}(x)$ is given by%
\[
\nu_{\gamma}(x)=\frac{1}{(1+x^{2})}\biggl[ \frac{1}{x^{1+\gamma
}}1\{
x\geq
0\}+\frac{1}{|x|^{1+\gamma}}1\{x<0\}\biggr].
\]
Due to   statements (i) and (ii) of Lemma \ref{MuProp}, one can
always choose $ \varepsilon$ in such a way that
$ \nu_{1} $ and $ \nu_{2} $ stay positive on $ \mathbb{R}_{+} $ and
thus they can be viewed as the L\'evy densities
of some L\'evy processes $L_{1,t}$ and $L_{2,t}$, respectively. It
directly follows from the definition of $ \nu_{1} $ and $ \nu_{2} $
that $ \nu_{1},\nu_{2}\in\mathfrak{B}_{\gamma}$. The next lemma
describes some other properties of $ \nu_{1}(x) $ and $ \nu_{2}(x)$.
Denote $ \bar\nu_{1}(x)=x^{2}\nu_{1}(x) $
and $ \bar\nu_{2}(x)=x^{2}\nu_{2}(x)$.
\begin{lem}
\label{NuProp}
Functions $ \bar\nu_{1}(x) $ and $ \bar\nu_{2}(x) $ satisfy
%
%
\begin{equation}
\label{NuNuh}
\sup_{x\in\mathbb{R}}|\bar\nu_{1}(x)-\bar\nu_{2}(x)|\geq
\varepsilon
Dh^{\beta}\log^{-1}(1/h)
\end{equation}
and
%
%
\begin{equation}
\label{FNu1Mom}
\int_{-\infty}^{\infty} (1+|u|^{\beta})\vert\mathbf{F}[\bar
{\nu}_{i
}](u)\vert\,du<\infty,\qquad i=1,2,
\end{equation}
that is, both functions $ \nu_{1}(x) $ and $ \nu_{2}(x) $ belong to the
class $ \mathfrak{S}_{\beta}$.
\end{lem}

Let us now perform a time change in the processes $ L_{1,t} $ and $
L_{2,t}$.
To this end, introduce a time change $ \mathcal{T}(t)$, such that
the Laplace transform of $%
\mathcal{T}(t)$ has following representation:%
\[
\mathcal{L}_{t}(z)=\E\bigl[e^{-z\mathcal{T}(t)}\bigr]=\int_{0}^{\infty
}e^{-zy}\,dF_{t}(y),
\]
where $ (F_{t}, t\geq0) $ is a family of distribution functions on
$ \mathbb{R}_{+} $ satisfying
\[
1-F_{t}(y)\leq1-F_{s}(y),\qquad y\in\mathbb{R}_{+},
\]
for any $ t\leq s$.
Denote by $ \widetilde{p}_{1,t} $ and $ \widetilde{p}_{2,t} $ the
marginal densities of the resulting time-changed L\'evy processes
$Y_{1,t}=L_{1,\mathcal{T}(t)}$ and $Y_{2,t}=
L_{2,\mathcal{T}(t)}$, respectively. The following lemma provides us
with an upper bound for the $ \chi^{2} $-divergence between
$\widetilde
{p}_{1,t}$ and $\widetilde{p}_{2,t}$, where for any two probability
measures $ P $ and $ Q $ the $ \chi^{2} $-divergence between $ P $ and
$ Q $ is defined as
\[
\chi^{2}(P,Q)=
\cases{\displaystyle
\int\biggl( \frac{dP}{dQ}-1 \biggr)^{2} \,dQ, &\quad if $P\ll Q$, \vspace*{2pt}\cr
+\infty, &\quad otherwise.}
\]
\begin{lem}
\label{BoundDiv}
Suppose that the Laplace transform of the time change $ \mathcal{T}(t)
$ fulfills
%
%
\begin{equation}
\label{LCond1}
\bigl\vert\mathcal{L}_{\Delta}^{(k+1)}(z)/\mathcal{L}_{\Delta
}^{(k)}(z)%
\bigr\vert=O(1),\qquad |z|\rightarrow\infty,
\end{equation}
for $ k=0,1,2$, and uniformly in $ \Delta\in[0,1]$. Then
\[
\chi^{2}( \widetilde{p}_{1,\Delta},\widetilde{p}_{2,\Delta
}
) \lesssim\Delta^{-1}[ \mathcal{L}_{\Delta}^{\prime
}(ch^{-\gamma
})%
] ^{2}h^{(2\beta+1)},\qquad h\rightarrow0,
\]
with some constant $ c>0$.
\end{lem}

The proofs of Lemmas \ref{MuProp}, \ref{NuProp} and \ref{BoundDiv} can
be found in the preprint version of our paper \citet{Bel}.
Combining Lemma \ref{BoundDiv} with  inequality~(\ref{NuNuh}) and
using the well-known Assouad  lemma [see, e.g., Theorem~2.6 in
\citet{TS}], one obtains
\[
\liminf_{n\to\infty}\inf_{\widehat\nu}\sup_{\nu\in\mathfrak
{B}_{\gamma}\cap\mathfrak{S}_{\beta}}\mathbb{P}\Bigl(\sup_{x\in\mathbb
{R}}|\bar\nu(x)-\widehat\nu(x) |>h^{\beta}_{n}\log
^{-1}(1/h_{n})\Bigr)>0
\]
for any sequence $ h_{n} $ satisfying
\[
n\Delta^{-1}[ \mathcal{L}_{t}^{\prime}(c\cdot h_{n}^{-\gamma})
] ^{2}h_{n}^{(2\beta+1)}=O(1),\qquad n\to\infty.
\]

\section{Auxiliary results}

\subsection{\texorpdfstring{Some results on time-changed L\'evy
processes}{Some results on time-changed Levy processes}}

\begin{lem}
\label{MIX}
Let $ L_{t} $ be a $ d $-dimensional L\'evy process with the L\'evy
measure $ \nu$ and let $ \mathcal{T}(t) $
be a time change independent of $ L_{t}$. Fix some $ \Delta>0 $ and
consider two sequences
$ T_{k}=\mathcal{T}(\Delta k)-\mathcal{T}(\Delta(k-1)) $ and $
Z_{k}=Y_{\Delta k}-Y_{\Delta(k-1)}$, $ k=1,\ldots, n$, where $
Y_{t}=L_{\mathcal{T}(t)}$. If the sequence $ (T_{k})_{k\in\mathbb{N}}
$ is strictly stationary and $ \alpha$-mixing with the mixing coefficients
$ (\alpha_{T}(j))_{j\in\mathbb{N}}$, then the sequence $
(Z_{k})_{k\in\mathbb{N}} $ is also strictly stationary and $ \alpha
$-mixing with the mixing coefficients $ (\alpha_{Z}(j))_{j\in\mathbb
{N}}$, satisfying
%
%
\begin{equation}
\label{AZAT}
\alpha_{Z}(j)\leq\alpha_{T}(j),\qquad j\in\mathbb{N}.
\end{equation}
\end{lem}
\begin{pf}
Fix some natural $ k,l $ with $ k+l<n$.
Using the independence of increments of the L\'evy process $ L_{t} $
and the fact that $ \mathcal{T} $
is a nondecreasing process, we get $ \E[ \phi(Z_{1},\ldots, Z_{k})
]=\E[ \widetilde\phi(T_{1},\ldots, T_{k}) ] $
and
\begin{eqnarray*}
&&\E[\phi(Z_{1},\ldots, Z_{k})\psi(Z_{k+l},\ldots, Z_{n})]\\
&&\qquad=\E
[\widetilde
\phi(T_{1},\ldots, T_{k}) \widetilde\psi(T_{k+l},\ldots, T_{n})],\qquad
k,l\in\mathbb{N},
\end{eqnarray*}
for any two functions $ \phi\dvtx\mathbb{R}^{k}\to[0,1] $ and $
\psi\dvtx
\mathbb{R}^{n-l-k}\to[0,1]$,
where $ \widetilde\phi(t_{1},\ldots,\allowbreak t_{k})=\E[\phi
(L_{t_{1}},\ldots,
L_{t_{k}})] $ and $ \widetilde\psi(t_{1},\ldots,t_{k})=\E[\psi
(L_{t_{1}},\ldots, L_{t_{k}})]$. This implies that
the sequence $ Z_{k} $ is strictly stationary and $ \alpha$-mixing
with the mixing coefficients satisfying
(\ref{AZAT}).
\end{pf}

\subsection{Exponential inequalities for dependent sequences}

The following theorem can be found in \citet{MPR}.
\begin{theorem}
\label{EIB}
Let $ (Z_k, k\geq1) $ be a strongly mixing sequence of centered
real-valued random variables on the probability space $(\Omega
,\mathcal F,P)$
with the mixing coefficients satisfying
%
%
\begin{equation}
\label{ALPHAEXPDECAY}
\alpha(n)\leq\bar\alpha\exp(-cn ),\qquad n\geq1, \bar\alpha
>0, c>0.
\end{equation}
Assume that $\sup_{k\geq1}|Z_k|\leq M$ a.s.,
then there is a positive constant $ C $ depending on $ c $ and $ \bar
\alpha$ such that
\[
\mathbb{P}\Biggl\{ \sum_{i=1}^n Z_i\geq\zeta\Biggr\}\leq\exp
\biggl[-\frac
{C\zeta^2 }{nv^{2}+M^{2} +M\zeta\log^{2}(n)}\biggr]
\]
for all $ \zeta>0 $ and $ n\geq4$,
where
\[
v^{2}=\sup_{i}\biggl( \E[Z_{i}]^{2}+2\sum_{j\geq i}\Cov(Z_{i},Z_{j})
\biggr).
\]
\end{theorem}
\begin{cor}
\label{COVEST}
Denote
\[
\rho_{j}=\E\bigl[ Z_{j}^{2}\log^{2(1+\varepsilon)}(
|Z_{j}|^{2}) \bigr],\qquad j=1,2,\ldots,
\]
with arbitrary small $ \varepsilon>0 $ and suppose that all $ \rho_{j}
$ are finite. Then
\[
\sum_{j\geq i}\Cov(Z_{i},Z_{j})\leq C\max_{j}\rho_{j}
\]
for some constant $ C>0$, provided (\ref{ALPHAEXPDECAY}) holds.
Consequently, the following inequality holds:
\[
v^{2}\leq\sup_{i}\E[Z_{i}]^{2}+C\max_{j}\rho_{j}.
\]
\end{cor}

The proof can be found in \citet{Bel}.
%
%
%

\subsection{Bounds on large deviations probabilities for weighted sup norms}
\label{EXPBOUNDS}

Let $ Z_{j}=(X_{j},Y_{j})$, $j=1,\ldots, n$, be a sequence of
two-dimensional random vectors and let $ G_{n}(u,z)$, $ n=1,2,\ldots, $
be a sequence of complex-valued functions defined on $ \mathbb{R}^{2}$. Define
\begin{eqnarray*}
\widehat m_{1}(u)&=&\frac{1}{n}\sum_{j=1}^{n}X_{j}G_{n}(u,X_{j}),
\\
\widehat m_{2}(u)&=&\frac{1}{n}\sum_{j=1}^{n}Y_{j}G_{n}(u,X_{j}),
\\
\widehat m_{3}(u)&=&\frac{1}{n}\sum_{j=1}^{n}X^{2}_{j}G_{n}(u,X_{j}),
\\
\widehat m_{4}(u)&=&\frac{1}{n}\sum_{j=1}^{n}X_{j}Y_{j}G_{n}(u,X_{j}).
\end{eqnarray*}
\begin{prop}
\label{ExpBounds}
Suppose that the following assumptions hold:
\begin{longlist}[(AZ1)]
\item[(AZ1)] The sequence $ Z_{j}$, $ j=1,\ldots, n$, is strictly
stationary and is $ \alpha$-mixing with mixing coefficients $ (\alpha
_{Z}(k))_{k\in\mathbb{N}} $ satisfying
\[
\alpha_{Z}(k)\leq\bar\alpha_{0}\exp(-\bar\alpha_{1} k),\qquad
k\in
\mathbb{N},
\]
for some $ \bar\alpha_{0}>0 $ and $ \bar\alpha_{1}>0$.
\item[(AZ2)] The r.v. $X_{j} $ and $ Y_{j} $ possess finite absolute
moments of order $ p>2$.
\item[(AG1)] Each function $ G_{n}(u,z), n\in\mathbb{N} $ is
Lipschitz in $ u $
with linearly growing (in~$ z $) Lipschitz constant,
that is, for any $ u_{1},u_{2} \in\mathbb{R} $
\[
|G_{n}(u_{1},z)-G_{n}(u_{2},z)|\leq L_{n}(a+b|z|)|u_{1}-u_{2}|,
\]
where $ a,b $ are two nonnegative real numbers
not depending on $ n $ and the sequence $ L_{n} $ does not depend on $
u$.
\item[(AG2)] There are two sequences $ \bar\mu_{n} $ and $ \bar
\sigma
_{n}$,
such that
\[
|G_{n}(u,z)|\leq\bar\mu_{n},\qquad (u,z)\in\mathbb{R}^{2},
\]
and all the functions
\begin{eqnarray*}
&\E[(|X|^{2}+|Y|^{2})|G_{n}(u,X)|^{2}],\qquad
\E[|X|^{4}|G_{n}(u,X)|^{2}],&\\
&\E[|X|^{2}|Y|^{2}|G_{n}(u,X)|^{2}]&
\end{eqnarray*}
are uniformly bounded on $ \mathbb{R} $ by $ \bar\sigma^{2}_{n}$.
Moreover, assume that the sequences $ \bar\mu_{n}, L_{n} $
and $ \bar\sigma_{n} $ fulfill
\begin{eqnarray*}
\bar\mu_{n}/\bar\sigma^{2}_{n}&=&O(1), \qquad\bar\mu_{n}/\bar
\sigma
_{n}=O(n^{1/2-\delta/2}),\qquad \bar\sigma^{2}_{n}=O(n),\\
L_{n}/\bar\sigma_{n}&=&O(n^{3/2}),\qquad n\to\infty,
\end{eqnarray*}
for some $ \delta$ satisfying $ 2/p<\delta\leq1$.\vadjust{\goodbreak}
\end{longlist}
Let $w$ be a symmetric, Lipschitz continuous, positive, monotone
decreasing on $\mathbb{R}_{+}$ function such that
%
%
\begin{equation}
\label{decreasingw}
0<w(z)\leq\log^{-1/2}(e+|z|),\qquad z\in\mathbb{R}.
\end{equation}
Then there is $ \delta'>0 $ and $ \xi_{0}>0 $, such that
the inequality
%
%
\begin{equation}
\label{MINEQ}
\mathbb{P}\Biggl\{\log^{-(1+\varepsilon)}(1+\bar\mu_{n})\sqrt{\frac
{n}{\bar\sigma
^{2}_{n}\log n}}
\| \widehat m_{k}- \E[\widehat m_{k}] \|_{L_{\infty
}(\mathbb
{R},w)}>\xi\Biggr\} \leq B n^{-1-\delta' }\hspace*{-28pt}
\end{equation}
holds for any $ \xi>\xi_{0}$, any $ k\in\{ 1,\ldots,4 \}$, some
positive constant $ B $ depending on $ \xi$ and arbitrary small $
\varepsilon>0$.
\end{prop}

The proof of the proposition can be found in \citet{Bel}.
\printaddresses

\end{document}